# Axisymmetric galaxy models with central black holes, with an application to M32

E.E. Qian,[1,2] P.T. de Zeeuw,[1] R.P. van der Marel[1,3] and C. Hunter[2]

[1] *Sterrewacht Leiden, Postbus 9513, 2300 RA Leiden, The Netherlands*
[2] *Department of Mathematics, Florida State University, Tallahassee FL 32306, USA*
[3] *Institute for Advanced Study, Princeton, NJ 08540, USA (present address)*

**ABSTRACT**

The contour integral method of Hunter & Qian is applied to axisymmetric galaxy models in which the distribution function (DF) is of the form $f = f(E, L_z)$, where $E$ and $L_z$ are the classical integrals of motion in an axisymmetric potential. A practical way to construct the unique even part $f_e(E, L_z)$ of the two–integral DF for such systems is presented. It is applied to models, both oblate and prolate, in which the mass density is stratified on similar concentric spheroids.

The spheroids with scale-free densities are discussed in detail. These provide useful approximations to the behaviour of more realistic models in the limit of small and large radii. The self-consistent case is treated, as well as the case in which there are additional contributions to the potential from a central black hole or dark halo. The two–integral DFs for scale-free densities in a Kepler potential are particularly simple. These can be used to model power–law density cusps near a central black hole, or to model the outer parts of finite-mass systems. The range of axis ratios and density profile slopes is determined for which spheroidal power-law cusps with a central black hole have a physical two-integral DF.

More generally, the two–integral DFs are discussed for a set of spheroidal '$(\alpha,\beta)$-models', characterized by a power-law density cusp with slope $\alpha$ at small radii, and a power-law density fall-off with slope $\alpha + 2\beta$ at large radii. As an application, the DF is constructed for the $(\alpha,\beta)$ model with a $1.8 \times 10^6\,\mathrm{M}_\odot$ black hole used by van der Marel et al. to interpret their high spatial resolution spectroscopic data for M32. The line-of-sight velocity profiles are calculated. The results confirm that the model fits the data remarkably well. The model is used to calculate the kinematic signatures of a central black hole in observations such as are now possible with the Hubble Space Telescope. The predicted Gaussian velocity dispersion for the M32 centre is $127\,\mathrm{km\,s}^{-1}$ with the $0.09'' \times 0.09''$ square aperture of the Faint Object Spectrograph, and $105\,\mathrm{km\,s}^{-1}$ with the $0.26''$ diameter circular aperture, while the central dispersion measured from ground-based data is only $86\,\mathrm{km\,s}^{-1}$.

## 1 INTRODUCTION

Few realistic dynamical models have been constructed for galaxies that are not spheres or discs. The main reason for this paucity of models is that in axisymmetric or triaxial galaxies the stellar motions are governed by non-classical integrals of motion, which are generally not known explicitly. An exception is provided by axisymmetric models in which the phase space distribution function (DF) $f = f(E, L_z)$, so that it depends only on the two classical integrals of motion, the energy $E$ and the angular momentum component $L_z$ parallel to the symmetry axis of the system. Hunter (1977) showed how the velocity dispersions in such a two–integral axisymmetric galaxy model can be calculated by solving the Jeans equations. Various authors have applied his solution to model kinematic observations of elliptical galaxies (e.g., Binney, Davies & Illingworth 1990; van der Marel, Binney & Davies 1990; van der Marel 1991; Cinzano & van der Marel 1994; Carollo & Danziger 1994). One disadvantage of this approach is that it is not evident whether the intrinsic velocity dispersions that best fit the line-of-sight measurements indeed correspond to a physical model, in which $f \geq 0$.

It is now possible to extract not only the mean line-of-sight velocity $\langle v_{\rm los}\rangle$ and velocity dispersion $\sigma_{\rm los}$ from absorp-tion line spectra, but also the shape of the line-of-sight velocity distribution, hereafter referred to as the *velocity profile* (VP) (e.g., Franx & Illingworth 1988; Bender 1990; Rix & White 1992; van der Marel & Franx 1993). To model such data one could continue to solve the Jeans equations of increasing order (e.g., Magorrian & Binney 1994), but it is preferable to calculate the entire DF, so that the theoretical VPs can be calculated accurately, and only models with $f \geq 0$ are considered.

The calculation of $f(E, L_z)$ for axisymmetric models has long been hampered by certain perceived technical difficulties, with as main result that only a few such DFs were found by various integral transform methods (e.g., Lynden-Bell 1962; Hunter 1975; Dejonghe 1986), usually for special mass models (but see Evans 1993, 1994; Evans & de Zeeuw 1994). Hunter & Qian (1993, hereafter HQ) showed how these difficulties can be circumvented, and developed a contour integral method that in principle allows calculation of $f(E, L_z)$ for a wide variety of mass models. In particular, it is no longer required that the density $\rho(R^2, z^2)$ can be written explicitly as a function of $\Psi$ and $R^2$, where $\Psi$ is the relative gravitational potential (cf. Binney & Tremaine 1987).

In this paper we demonstrate how the HQ method can be used to calculate $f(E, L_z)$ for realistic axisymmetric galaxy models, with emphasis on models stratified on similar concentric spheroids with arbitrary density profiles. As a specific application, we study a set of '$(\alpha, \beta)$-models', characterized by a power-law density cusp with slope $\alpha$ at small radii, and a power-law density fall-off with slope $\alpha + 2\beta$ at large radii. The HQ method allows us to include an external potential, such as that of a central black hole or a dark halo. The present paper complements recent work by Dehnen & Gerhard (1994) who consider two-integral axisymmetric models with central density cusps, based on a convenient series expansion of $f(E, L_z)$, and give a thorough discussion of the VPs. We derive the DFs for a larger class of models, include the effects of a central black hole or dark halo, and derive some further properties of the VPs. In addition, we present a detailed study of the spheroids with scale-free mass densities, and show how they can be used to approximate the dynamical structure of the more general models at small and large radii.

We illustrate our technique by applying it to the nearby elliptical galaxy M32, which may contain a black hole (or at least a dark mass concentration) in its centre. Van der Marel et al. (1994b) used axisymmetric two-integral models to interpret the high spatial resolution kinematic observations of M32 by van der Marel et al. (1994a). The modelling consisted of: (i) use of Evans' (1994) power-law model DFs without a central black hole; and (ii) calculation of the first three moments of the VP for the case with a black hole, by solution of the moment equations of the collisionless Boltzmann equation. A remarkably good fit was obtained with a $1.8 \times 10^6$ M$_\odot$ black hole, but the actual DF could not be calculated. With the technique presented here we *can* calculate the entire $f(E, L_z)$ for the model with a central black hole, allowing a better comparison with the available data. We confirm and strengthen the results of van der Marel et al., and use the DF to calculate the kinematics and VP shapes that one would expect to observe with the high spatial resolution of the Hubble Space Telescope (HST).

This paper is organized as follows. In Section 2 we discuss the implementation of the contour integral method to cases where the relation $\rho = \tilde{\rho}(\Psi, R^2)$ is known implicitly, show how for similar concentric spheroids $f(E, L_z)$ can be found as a numerical quadrature for each $E$ and $L_z$, and summarize how the VPs of two-integral axisymmetric models can be calculated. In Section 3 we describe the properties of a family of spheroidal mass models with a central density cusp. We consider the special case of scale-free spheroidal models in detail, and discuss the inclusion of a central black hole. We apply the results to M32 in Section 4, and summarize our conclusions in Section 5.

## 2 TWO-INTEGRAL DISTRIBUTION FUNCTIONS

A general algorithm for the application of the HQ method to models in which the density as a function of potential and cylindrical radius is only known implicitly is presented in Section 2.1. The case of spheroidal mass models with arbitrary density profiles is discussed in Section 2.2. Section 2.3 addresses the calculation of VPs for two-integral DFs. The reader who is interested mainly in the applications of the method can skip to Section 3.

### 2.1 The HQ contour integral method

We consider an axisymmetric model of infinite extent with a density $\rho(R^2, z^2)$ and an overall potential $\Psi(R^2, z^2)$. In a self-consistent system $\rho$ and $\Psi$ are related via Poisson's equation while in a non-self-consistent system $\Psi$ contains contributions from other components, which may include a dark halo and/or a central black hole, besides that from the density $\rho$. The HQ method is applicable in both cases. We adopt the convention in which the potential attains its maximum value at the centre and decreases outwards. Hence $z^2$ is determined uniquely by the values of $\Psi$ and $R^2$, provided that the former lies between the potential at infinity $\Psi_\infty$ and the equatorial potential $\Psi(R^2, 0)$. Therefore the density as a function of $\Psi$ and $R^2$, which we denote as $\tilde{\rho}(\Psi, R^2)$, can be obtained. This function, whose analytic continuation is needed in the HQ method, is implicit in cases where $z^2$ can only be determined implicitly for a given pair $(\Psi, R^2)$. It is this implicit case of the contour integral method with which we shall be concerned.

The HQ method can be used for both finite and infinite mass systems. When applied to the density $\rho(R^2, z^2)$ in a potential $\Psi(R^2, z^2)$ it gives the unique $f_e(E, L_z)$ that is *even* in $L_z$ and that generates $\rho$. When applied to $R\rho\langle v_\phi\rangle(R^2, z^2)$ it gives the unique $f_o(E, L_z)$ that is *odd* in $L_z$ and generates the mean azimuthal streaming motions $\langle v_\phi\rangle$. In practice it is not easy to obtain good observational data on the two-dimensional mean $\langle v_{\text{los}}\rangle$ on the plane of the sky from which the intrinsic azimuthal mean streaming field $\langle v_\phi\rangle(R, z)$ must be found. Therefore, an alternative approach is to take the odd part $f_o$ as a product of the even part $f_e$ and a prescribed function whose magnitude is no greater than unity. This ensures that $f = f_e + f_o$ is physical (non-negative) if $f_e$ is.

The physical values of $E$ and $L_z$ are those which correspond to bound orbits in the potential $\Psi$. They lie in the region of the $(E, L_z^2)$-plane that is bounded by the two straight lines $L_z = 0$ and $E = \Psi_\infty$, and by the curve $\mathcal{E}$, as shown in Figure 1. When $\Psi_\infty = -\infty$, this region extends indefinitely to the left. The curve $\mathcal{E}$ is the locus of the circular orbits in the equatorial plane. It is described by the parametric relation

$$E = \Psi(R_c^2, 0) + R_c^2 \frac{d\Psi(R^2, 0)}{dR^2}\bigg|_{R=R_c},$$
$$L_z^2 = -2R_c^4 \frac{d\Psi(R^2, 0)}{dR^2}\bigg|_{R=R_c}, \quad (2.1)$$

where $R_c$ is the radius of the circular orbit. The value $L_c \equiv L_z(R_c)$ is the maximum allowed value of $L_z$ at fixed energy. The set of orbits with energy $E$ and angular momentum $L_z$ cover a range $[\Psi_{\min}, \Psi_{\max}]$ of physically achievable values of the potential energy $\Psi$. This range can be found geometrically by constructing the two straight lines through the point $(E, L_z^2)$ that are tangent to the locus $\mathcal{E}$ in the Lindblad diagram (Figure 1). Their respective intersections with the boundary $L_z = 0$ give $\Psi_{\max}$ and $\Psi_{\min}$, and so mark a *window* $\mathcal{P}$ on this line (HQ). We denote the potential energy of the circular orbit of energy $E$ by $\Psi_{\mathrm{env}}(E)$. Its value can be found geometrically (Figure 1), and it also follows easily upon solution of the first of equations (2.1) for $R_c$, and then using $\Psi_{\mathrm{env}}(E) = \Psi(R_c^2, 0)$. The window $\mathcal{P}$ and the value $\Psi_{\mathrm{env}}(E)$ in it are important in the evaluation of $f(E, L_z)$ by means of the HQ method.

**Figure 1.** The physical domain of bound orbits in the $(E, L_z^2)$-plane (Lindblad diagram) is bounded by $L_z = 0$, $E = \Psi_\infty$ and the locus $\mathcal{E}$ of circular orbits in the equatorial plane, defined by equation (2.1). The thin straight lines through the point $(E, L_z^2)$ indicated by the solid dot are tangent to $\mathcal{E}$, and intersect the boundary $L_z = 0$ at values $\Psi_{\min}$ and $\Psi_{\max}$ which bound the window $\mathcal{P}$ of physically achievable values of the potential energy $\Psi$ for orbits with integrals $E$ and $L_z$. The special value $\Psi_{\mathrm{env}}$ in $\mathcal{P}$ is the intersection of the straight line that is tangent to $\mathcal{E}$ at energy $E$ and $L_z$ equal to the maximum allowed value $L_c$.

One way of writing the HQ solution for the even part $f_e(E, L_z)$ of the DF is

$$f_e(E, L_z) =$$
$$\frac{1}{4\pi^2 \mathrm{i}\sqrt{2}} \int_{\Psi_\infty}^{[\Psi_{\mathrm{env}}(E)+]} \tilde{\rho}_{11}\left[\xi, \frac{L_z^2}{2(\xi - E)}\right] \frac{d\xi}{(\xi - E)^{1/2}}. \quad (2.2)$$

This is a complex contour integral on the complex $\xi$-plane. The "density" term of the integrand is now a function of the single complex variable $\xi$ (the complex potential), and the two subscripts 1 denote the *second* partial derivative with respect to its *first* argument (as in HQ). The value of this function in the complex domain is obtained via the analytic continuation of the physically relevant value $\tilde{\rho}_{11}(\Psi, R^2)$. For simplicity we also denote the analytic continuation by $\tilde{\rho}_{11}$. The physically achievable values of $\Psi$ lie on the real $\xi$ axis to the right of the point $\xi = E$ in the window $\mathcal{P}$, i.e., for $\xi = \Psi$ in $\mathcal{P}$, the point $[\Psi, R^2 = \frac{1}{2}L_z^2/(\Psi - E)]$ lies in the physical domain of $\tilde{\rho}_{11}$. Obviously values of $\tilde{\rho}_{11}$ on the window $\mathcal{P}$ must coincide with the physically relevant values there. For the square root term in the integrand we choose the branch induced by a cut to the left of the point $\xi = E$ along the real $\xi$-axis so that it is real and positive when $\xi > E$. This choice together with the fact that $\mathcal{P}$ is to the right of $\xi = E$ ensures that the integrand is real for $\xi$ in $\mathcal{P}$.

The point $\xi = \Psi_{\mathrm{env}}(E)$, which only depends on $E$, is a point that always lies in the window $\mathcal{P}$. As indicated by the notation $[\Psi_{\mathrm{env}}(E)+]$ in equation (2.2), the path of the complex contour integral is taken as a loop which starts on the lower side of the real $\xi$-axis at $\xi = \Psi_\infty$, crosses the real $\xi$-axis at $\xi = \Psi_{\mathrm{env}}(E)$, and ends at $\xi = \Psi_\infty$ on the upper side of the real $\xi$-axis. To evaluate the integral (2.2) we first specify a contour. Depending on whether $\Psi_\infty$ is finite or (negative) infinite, a different parametrization of the contour must be used (Figure 2). For $\Psi_\infty$ finite, one can always choose $\Psi_\infty = 0$. A simple parametrization which corresponds to an elliptical path is given by the equation

$$\xi = \tfrac{1}{2}\Psi_{\mathrm{env}}(E)(1 + \cos\theta) + \mathrm{i}\, h \sin\theta, \quad -\pi \leq \theta \leq \pi. \quad (2.3)$$

For $\Psi_\infty = -\infty$ we can choose

$$\xi = \Psi_{\mathrm{env}}(E) + l\left(1 - \sec\frac{\theta}{2}\right) + \mathrm{i}\, h \sin\theta, \quad -\pi \leq \theta \leq \pi. \quad (2.4)$$

The parameter $h$ determines the maximum width of the contour in both equations (2.3) and (2.4). It must be kept small in order for the contour to avoid enclosing any complex conjugate singularities that the "density" term $\tilde{\rho}_{11}$ might have. However too small a value of $h$ will force the contour to come near enclosed singularities on the real $\xi$-axis, when greater care must be taken to achieve good accuracy of the numerical integration. The parameter $l$ in equation (2.4) allows the location of the points of maximum width of the contour at $\theta = \pm \pi/2$ to be adjusted. We take both $h$ and $l$ to be of the order of $0.1\Psi_{\mathrm{env}}(E)$. These parametrizations convert the complex contour integral (2.2) into an integration with respect to the angle $\theta$, and have been satisfactory in our computations. Other parametrizations are possible though. For $\Psi_\infty = -\infty$ it is sometimes convenient to first change the integration variable in the solution (2.2), to obtain an integral with a finite path (see Appendix B).

The fact that the integrand in equation (2.2) is real-valued on $\mathcal{P}$ has a useful consequence. According to the Schwarz Reflection Principle (Levinson & Redheffer 1970), once we have succeeded in continuing the integrand into a domain above the real $\xi$-axis, we can also continue it analytically as a complex conjugate into the reflected domain below the real $\xi$-axis. Therefore we can evaluate $f_e$ by integrating along either the upper or the lower half of the loop and multiplying the result by a factor of 2. For definiteness

**Figure 2.** The complex $\xi$-plane, with the contour used in the numerical evaluation of $f(E, L_z)$ by means of equation (2.2). Top panel: the contour (2.3), for the case when $\Psi_\infty = 0$. Bottom panel: the contour (2.4) for the case when $\Psi_\infty = -\infty$. The window $\mathcal{P}$ of physically allowed values of $\Psi$ is indicated by the thick solid bar along the real $\xi$-axis (see Figure 1). The contour intersects this window at the value $\Psi_{\rm env}(E)$, which is the potential energy of the circular orbit with energy $E$.

we shall use the upper half for our calculations.

To evaluate the integral (2.2) numerically for a given pair $(E, L_z)$, we first discretize the contour (2.3) or (2.4) by a Gauss quadrature in $\theta$, and then approximate the integral by a weighted sum of the values of the integrand at the quadrature points. The main task is then to evaluate $\tilde{\rho}_{11}[\xi, R^2 = \frac{1}{2}L_z^2/(\xi - E)]$ at each quadrature point $\xi$ on the contour. By implicit differentiation, we obtain

$$\tilde{\rho}_{11}(\xi, R^2) = \frac{\rho_{22}(R^2, z^2)}{[\Psi_2(R^2, z^2)]^2} - \frac{\rho_2(R^2, z^2)\Psi_{22}(R^2, z^2)}{[\Psi_2(R^2, z^2)]^3}, \quad (2.5)$$

in which *each* subscript 2 denotes a partial differentiation with respect to $z^2$. In using this equation, we only have to find the value of $z^2$ for a given pair $[\xi, R^2 = \frac{1}{2}L_z^2/(\xi - E)]$. For the case under consideration, a complex equation has to be solved at each quadrature point. We emphasize that the contour integral solution requires that the integrand attains its physically achieved values in the window $\mathcal{P}$ and its values on the complex contour are from the analytic continuation of the density. Therefore it is absolutely essential that the values of $z^2$, which are to be obtained from the numerical solution of the equation

$$\xi = \Psi\left[\frac{L_z^2}{2(\xi - E)}, z^2\right], \quad (2.6)$$

satisfy this requirement.

We thus proceed as follows. We choose a pair $(E, L_z)$, and start at the point $\xi = \Psi_{\rm env}(E)$ ($\theta = 0$) in the window $\mathcal{P}$. Since we are now in the physical domain, we look for the unique real positive solution of $z^2$ of equation (2.6). As we move along the contour and arrive at a quadrature point, we use the $z^2$ value at the previous point as the initial guess and employ Newton's method to solve equation (2.6) iteratively. Numerical calculations we have done so far show that this approach provides close initial guesses for $z^2$ so that the iterations quickly converge to the correct branch. Once $z^2$ is found we use equation (2.5) to evaluate the integrand, which is then added to the weighted sum for the Gauss quadrature. As we approach $\xi = \Psi_\infty$ along the contour, $R^2 = \frac{1}{2}L_z^2/(\xi - E)$ approaches a finite value, hence $z^2$ becomes unbounded. Consequently any numerical method that we use for equation (2.6) breaks down. However the "density" term $\tilde{\rho}_{11}$ is vanishingly small at large distances for centrally condensed systems, so that its contribution to the integral (2.2) becomes negligible. This can be measured by the ratio of the value of the integrand to the value of the weighted sum at each given quadrature point. Once this ratio falls below a preset tolerance we stop and accept the weighted sum as the value of the integral (2.2).

### 2.2 Classical spheroids

We now restrict our attention to axisymmetric systems of infinite extent in which the density of stars is stratified on similar concentric spheroids, i.e.

$$\rho = \rho(m^2), \qquad m^2 = R^2 + z^2/q^2, \quad (2.7)$$

and $q$ is the axis ratio. Oblate models have $0 < q < 1$, and prolate models have $q > 1$. We write the overall potential as a sum $\Psi = \Psi^* + \Psi^{\rm ext}$, in which the first term is the potential induced by the stellar density (2.7) and the second represents contributions by external components such as a central black hole or a dark halo. Different mass densities $\rho(m^2)$ require different expressions for the associated potential $\Psi^*$. The classical theory of the gravitational potential of ellipsoidal bodies (e.g., Chandrasekhar 1969) gives two alternative expressions:

$$\Psi^*(R^2, z^2) = \pi G q \int_0^\infty \frac{du}{\Delta(u)} \int_U^\infty \rho(m^2)\, dm^2,$$

$$= \Psi_0^* - \pi G q \int_0^\infty \frac{du}{\Delta(u)} \int_0^U \rho(m^2)\, dm^2, \quad (2.8)$$

where $G$ is the gravitational constant, $\Delta(u)$ and $U$ are defined as

$$\Delta(u) = (1 + u)\sqrt{q^2 + u},$$
$$U = \frac{R^2}{1 + u} + \frac{z^2}{q^2 + u}, \quad (2.9)$$

and $\Psi_0^*$ is the central potential

$$\Psi_0^* = \frac{2\pi G q \arcsin e}{e} \int_0^\infty \rho(m^2)\, dm^2, \quad (2.10)$$

with $e = \sqrt{1 - q^2}$. In prolate models $e$ is imaginary, but $e^{-1}\arcsin e$ is real, and equals $(q^2 - 1)^{-1/2}\mathrm{arcsinh}(q^2 - 1)^{1/2}$. The two expressions in equation (2.8) are equivalent only when the integral (2.10) converges, and the potential is finite everywhere. However, we also need to consider two ways in which this integral may diverge. It may diverge only at the lower limit $m^2 = 0$ due to a strong central cusp. In this case the potential can be taken as the first expression in equation (2.8); it is positive infinite at the centre and vanishes at large distances. Alternatively, the integral may diverge only at its upper limit $m^2 = \infty$. Then the potential can be taken as the second expression of equation (2.8), in which $\Psi_0^*$ now is just an additive constant. In this case the potential has the finite value $\Psi_0^*$ at the centre and becomes negative infinite at large distances. When the integral in equation (2.10) diverges at both its lower and upper limits, neither expression given in equation (2.8) is applicable, since both inner integrals now diverge. We must then replace the fixed limits of the inner integrals by some interior value of $m^2$ and so take a finite part of these divergent integrals. The resulting potential is positive infinite at the centre and negative infinite at large distances. When convenient, as it is in Section 3.2 below, a constant can be added to the potential in all cases.

The double integration (2.8) can be carried out explicitly in some special cases. More often, only the inner integration can be done analytically, and a one-dimensional outer integral remains. Some examples are given in Section 3. It is always possible to exchange the order of the integration in equation (2.8) to reduce it to a one-dimensional integral. While this exchange is simple for $R^2$ and $z^2$ in the physical range (i.e., both non-negative), it must be done more carefully for the wider range of values of $R^2$ and $z^2$, which includes complex ones, on which our contour integral method operates. For complex values of $R^2$ and $z^2$ we proceed as follows. Assuming that $U = R^2/(1+u) + z^2/(q^2+u)$ lies in the region in which $\rho(m^2)$ is analytic, we let $m^2 = R^2/(1+x) + z^2/(q^2+x)$, $x \in [u, \infty)$ be the path for the inner complex integral of the second expression of equation (2.8). This substitution converts the inner integral into one with respect to the real variable $x$. Then an exchange of the order of integration followed by a simple integration yields a one-dimensional integral. The result is

$$\Psi^*(R^2, z^2) = \frac{2\pi Gq}{e}\left\{\arcsin e \int_{R^2+\frac{z^2}{q^2}}^{\infty} \rho(m^2)\,dm^2 + \int_0^\infty \rho(U)\left[\frac{R^2}{(1+u)^2} + \frac{z^2}{(q^2+u)^2}\right]\arcsin\frac{e}{\sqrt{1+u}}\,du\right\},$$

$$= \Psi_0^* - \frac{2\pi Gq}{e}\int_0^\infty \rho(U)\left[\frac{R^2}{(1+u)^2} + \frac{z^2}{(q^2+u)^2}\right] \times \left(\arcsin e - \arcsin\frac{e}{\sqrt{1+u}}\right)du. \quad (2.11)$$

This allows evaluation of $\Psi$ by a one-dimensional (numerical) quadrature even for complex values of $R^2$ and $z^2$.

The partial derivatives needed in equation (2.5) can now also be calculated easily. We find

$$\Psi_2 = -\pi Gq\int_0^\infty \frac{du}{\Delta(u)(q^2+u)}\rho(U) + \Psi_2^{\text{ext}},$$

$$\Psi_{22} = -\pi Gq\int_0^\infty \frac{du}{\Delta(u)(q^2+u)^2}\rho'(U) + \Psi_{22}^{\text{ext}}, \quad (2.12)$$

$$\rho_2 = \rho'(m^2)q^{-2}, \qquad \rho_{22} = \rho''(m^2)q^{-4}.$$

In order to add a central black hole with mass $M_{\text{BH}}$, we take $\Psi^{\text{ext}} = GM_{\text{BH}}/\sqrt{R^2 + z^2}$.

With these formulas in place, we can choose a pair $(E, L_z)$ in the physical domain (Figure 1), and then at each quadrature point $\xi$ evaluate $R^2 = \frac{1}{2}L_z^2/(\xi - E)$, solve equation (2.6) for $z^2$ according to the procedure given in Section 2.1, evaluate equation (2.5), and then compute the contribution to $f_e(E, L_z)$ at the point $\xi$.

### 2.3 Velocity profiles

The observable properties of the two-integral axisymmetric models include the line-of-sight velocity moments (e.g., the mean streaming velocity $\langle v_{\text{los}}\rangle$ and the velocity dispersion $\sigma_{\text{los}}$, defined as $\sigma_{\text{los}}^2 = \langle v_{\text{los}}^2\rangle - \langle v_{\text{los}}\rangle^2$), and the entire VP shape. The intrinsic velocity dispersions can be calculated conveniently by direct integration of the Jeans equations (Hunter 1977; Appendix C). Integration along the line of sight can then be done using the expressions given by, e.g., Evans & de Zeeuw (1994). The higher order moments can be found in a similar way (e.g., Magorrian & Binney 1994). Here we discuss only the calculation of the observed VP.

We let $(x, y, z)$ be Cartesian coordinates with the $z$-axis the symmetry axis of the model. We use $(x', y', z')$ as the Cartesian coordinates of an observer, where the $z'$-axis lies along the line of sight, and the $x'$ and $y'$ axes are oriented along the major and minor axes of the projected surface density of the galaxy. We assume that the galaxy is seen at an inclination angle $i$. Then $x = -y'\cos i + z'\sin i$, $y = x'$ and $z = y'\sin i + z'\cos i$.

The normalized VP of an axisymmetric $f(E, L_z)$ galaxy is

$$\text{VP}(v_{z'}; x', y') = \frac{1}{\Sigma}\int\int\int_{E > \Psi_\infty} f(E, L_z)\,dv_{x'}\,dv_{y'}\,dz', \quad (2.13)$$

where $v_{x'}$ and $v_{y'}$ are two velocity components on the plane of the sky, $v_{z'}$ is the line-of-sight velocity $v_{\text{los}}$, and $\Sigma(x', y') = \int \rho dz'$ is the projected surface brightness. We employ polar coordinates $(v_\perp, \varphi)$ in the $(v_{x'}, v_{y'})$-velocity space, with $v_{x'} = v_\perp\cos\varphi$ and $v_{y'} = v_\perp\sin\varphi$. Then

$$\text{VP}(v_{z'}; x', y') = \frac{1}{\Sigma}\int_{z_1}^{z_2} dz' \int_0^{2(\Psi - \Psi_\infty) - v_{z'}^2} dv_\perp^2 \int_0^\pi f(E, L_z)\,d\varphi, \quad (2.14)$$

where

$$E = \Psi(x', y', z') - \frac{1}{2}(v_{z'}^2 + v_\perp^2),$$
$$L_z = -v_{z'}x'\sin i \quad (2.15)$$
$$+ v_\perp\cos\varphi\,\sqrt{(-y'\cos i + z'\sin i)^2 + x'^2\cos^2 i}.$$

Here $\Psi = \Psi(x', y', z')$ is the potential expressed in observer's coordinates. The term $\cos\varphi$ should contain a phase shift $\varphi_0$, but because of the periodicity of the cosine function, and the integration over the full range of $\varphi$, we can neglect $\varphi_0$, set the range of the $\varphi$-integration to $(0, \pi)$, and multiply by two. For cases with $\Psi_\infty = 0$, the range of $v_{z'}$ is finite and confined by the maximum escape velocity along the line of sight, and $z_1$ and $z_2$ are two extreme points along the line of sight determined by $\Psi(x', y', z') - \frac{1}{2}v_{z'}^2 = 0$. If $\Psi_\infty = -\infty$, the range of $v_{z'}$ extends from $-\infty$ to $\infty$ and so does the range of $z'$. Hence $z_1 = -\infty$ and $z_2 = \infty$ in this case. We shall generally evaluate the triple integral (2.14) by numerical quadrature.

The *even* and *odd* parts of the VP are defined as

$$\mathrm{VP}_{e,o}(v_{z'}; x', y') = \tfrac{1}{2}[\mathrm{VP}(v_{z'}; x', y') \pm \mathrm{VP}(-v_{z'}; x', y')]. \quad (2.16)$$

According to the above analysis, $\mathrm{VP}(-v_{z'}; x', y')$ is given by equation (2.14) with only a sign change of the first term of the angular momentum in equation (2.15). Since $\int_0^\pi F(\cos\varphi)d\varphi = \int_0^\pi F(-\cos\varphi)d\varphi$ for an arbitrary function $F$, we can switch the sign of the second term of the angular momentum in equation (2.15), which then becomes $-L_z$. This relation hence allows us to establish the correspondence

$$\mathrm{VP}_{e,o}(v_{z'}; x', y') = \frac{1}{\Sigma} \int_{z_1}^{z_2} dz' \int_0^{2(\Psi-\Psi_\infty)-v_{z'}^2} dv_\perp^2 \int_0^\pi f_{e,o}(E, L_z) d\varphi, (2.17)$$

The integral (2.14) for the VP simplifies when the term $-v_{z'}x' \sin i$ in expression (2.15) for $L_z$ vanishes. We show in Appendix A that this allows a reduction to a straightforward integration over the density distribution itself. This is useful for the calculation of three special cases: (i) the VP on the minor axis ($x' = 0$) for arbitrary inclination, (ii) the VP of a face-on galaxy ($i = 0$), and (iii) the density distribution of stars which have the systemic velocity, $v_{z'} = 0$.

## 3 SPHEROIDAL MODELS

We now consider a specific family of models with $\rho(m^2)$. An application of these spheroidal models to the galaxy M32 is discussed in Section 4.

### 3.1 The $(\alpha, \beta)$-models

A large number of models with $\rho = \rho(m^2)$ have been used in dynamical studies of galaxies (e.g., Perek 1962; de Zeeuw & Pfenniger 1988). We consider the family of models defined by

$$\rho(m^2) = \rho_0 \left(\frac{m}{b}\right)^\alpha \left(1 + \frac{m^2}{b^2}\right)^\beta, \quad (3.1)$$

where the exponents $-3 < \alpha \leq 0$ and $\beta \leq 0$ are parameters, $\rho_0$ is a reference density, and $b$ a reference length. When $\alpha = 0$, the central density is finite and equal to $\rho_0$. The density profile has a central cusp when $\alpha < 0$. When $\beta = 0$, the models are scale-free. At large radii the density falls off proportional to $r^{\alpha+2\beta}$.

When observed at inclination $i$, these spheroidal models have a projected surface density $\Sigma = \Sigma(m'^2)$, given by

$$\Sigma(m'^2) = \frac{q}{q'} \int_{m'^2}^{\infty} \frac{\rho(m^2) \, dm^2}{\sqrt{m^2 - m'^2}}, \quad (3.2)$$

where $m'^2 = x'^2 + y'^2/q'^2$, so that the isophotes are similar concentric ellipses with an observed axis ratio $q'$ which is given by

$$q'^2 = \cos^2 i + q^2 \sin^2 i. \quad (3.3)$$

For the models defined in equation (3.1), $\Sigma$ falls off proportional to $(R')^{\alpha+2\beta+1}$ at large projected radii $R'$. It has a central power-law cusp for $\alpha < -1$. When $\alpha = -1$ this cusp is logarithmic. The central projected surface density is finite when $\alpha > -1$.

Figure 3 shows a diagram of the $(\alpha, \beta)$-parameter space. It is divided into regions by the straight lines $\alpha$=constant and $\alpha + 2\beta$=constant. When $\alpha > -2$ and $\alpha + 2\beta < -2$ (horizontally hatched region), the potential assumes either expression of equation (2.8) and its central value is given by

$$\Psi_0^* = \frac{2\pi G q \rho_0 b^2 \arcsin e}{e} B(\tfrac{\alpha}{2}+1, -\tfrac{\alpha}{2}-\beta-1), \quad (3.4)$$

where $B$ is the beta function. To the left of this region ($\alpha \leq -2$) the potential is given by the first form of equation (2.8), and to the right ($\alpha + 2\beta \geq -2$) by the second form. When $(\alpha, \beta)$ lies in the vertically hatched region bounded by the lines $\alpha = -3$ and $\alpha + 2\beta = -3$, the system has finite total mass

$$M = 2\pi q \rho_0 b^3 B(\tfrac{\alpha}{2}+\tfrac{3}{2}, -\tfrac{\alpha}{2}-\beta-\tfrac{3}{2}). \quad (3.5)$$

The line $\alpha = -\tfrac{1}{2}$ is significant, since a two-integral model with a central black hole is physical ($f \geq 0$) only when $\alpha \leq -\tfrac{1}{2}$ (Section 3.3).

Some special cases of the family (3.1) have been used in dynamical studies before. Along the right boundary of Figure 3, models with $\alpha = 0$ and integer or half integer values of $\beta$ are of interest since their potentials are either elementary or can be given in terms of special functions (de Zeeuw & Pfenniger 1988). The $(\alpha, \beta) = (0, -2)$ models are the perfect spheroids (Kuzmin 1956; de Zeeuw 1985) which admit three integrals of motion. The scale-free spheroids lie along the top boundary ($\beta = 0$). They are attractive candidates for detailed investigation for two reasons. One is that their internal dynamics is simpler than general models and the other is that they provide good approximations to the inner region of cusped models and to the outer region of a wide range of spheroidal models. We refer to the models with $(\alpha, \beta) = (-2, 0)$ as the *singular isothermal spheroids*, since they are the flattened counterparts of the well-known singular isothermal sphere. The solid line that connects $(\alpha, \beta) = (-2, -1)$ to $(\alpha, \beta) = (0, -2)$ indicates a set of models that are very similar to ones studied recently by Dehnen & Gerhard (1994). Their model densities are like equation (3.1), but with $(1 + m/b)^{2\beta}$ as second term rather than $(1 + m^2/b^2)^\beta$.

The function $\tilde{\rho}(\Psi, R^2)$ generally can not be given explicitly for models with $\rho = \rho(m^2)$. This is true also for the $(\alpha, \beta)$-family, even in cases where the potential is elementary. The two-integral DFs can be found by means of the method described in Section 2. The calculations simplify for the limiting case of the scale-free spheroids ($\beta = 0$), which we discuss below.

**Figure 3.** The $(\alpha,\beta)$ parameter space that governs the properties of the mass models defined in equation (3.1). Scale–free models have $\beta = 0$. Models in the vertically hatched area have finite total mass. The central potential is finite in the horizontally hatched area. Models with a central black hole have a physical $f_e(E, L_z)$ when $\alpha \leq 0.5$, i.e., to the left of the dashed vertical line. The dots at $(\alpha,\beta) = (-2,0)$ and $(0,-2)$ indicate, respectively, the singular isothermal spheroids and the perfect spheroids. The filled square at $(\alpha,\beta) = (-2,-1)$ indicates models with a Jaffe (1983) like profile. The solid line which connects it to the perfect spheroids indicates the set of models that is nearly identical to the family studied by Dehnen & Gerhard (1994). The asterisk indicates the values of $\alpha$ and $\beta$ appropriate for the galaxy M32, discussed in Section 4.

### 3.2 Scale–free spheroids

The density distribution of the scale–free spheroids can be written as

$$\rho = \rho_0(\bar{R}^2 + \bar{z}^2/q^2)^{\alpha/2} = \rho_0 \bar{r}^\alpha (\sin^2\theta + q^{-2}\cos^2\theta)^{\alpha/2}, \quad (3.6)$$

where $\bar{R} = R/b$ and $\bar{z} = z/b$ are dimensionless variables, and $(\bar{r},\theta)$ are scaled polar coordinates defined by $\bar{R} = \bar{r}\sin\theta$ and $\bar{z} = \bar{r}\cos\theta$. This shows that the density is a product of a power of the radius times a function of $\theta$. The total mass of these spheroids is infinite. The projected surface density is $\Sigma = \Sigma_0(x'^2 + y'^2/q'^2)^{(1+\alpha)/2}$, with $q'$ given in equation (3.3) and $\Sigma_0 = \rho_0 b^{-\alpha} B(\frac{1}{2}, -\frac{\alpha}{2} - \frac{1}{2}) q/q'$.

The potential $\Psi_0^*$ of equation (2.10) diverges for all scale-free spheroids. We therefore replace the fixed limit of the inner integral in equation (2.8) by some interior value, and make use of the flexibility to add a convenient constant. We take the gravitational potential of the singular isothermal spheroid ($\alpha = -2$) to be

$$\Psi = -\pi G q \rho_0 b^2 \int_0^\infty \frac{du}{\Delta(u)} \ln\left[\bar{R}^2 + \bar{z}^2\left(\frac{1+u}{q^2+u}\right)\right]. \quad (3.7)$$

It is $\infty$ at the centre and $-\infty$ at large radii. It can not be expressed in terms of elementary functions, but the associated forces can (de Zeeuw & Pfenniger 1988). We take the potentials of the other scale–free spheroids to be

$$\Psi = -\frac{2\pi G q \rho_0 b^2}{(\alpha+2)} \int_0^\infty \frac{du}{\Delta(u)} \left[\bar{U}^{\alpha/2+1} - (1+u)^{-(\alpha/2+1)}\right], \quad (3.8)$$

where $\bar{U} = U/b^2$ and $U$ is defined in equation (2.9). The additive constant given by the second term in the bracket in expression (3.8) is

$$\Psi_c = V_0^2/(\alpha+2), \quad (3.9)$$

where we have written

$$V_0^2 = 2\pi G q \rho_0 b^2 J_{\alpha,q}, \quad (3.10)$$

and $J_{\alpha,q}$ is the auxiliary integral defined in equation (B3). The choice of the additive constant $\Psi_c$ ensures that the $\alpha \to -2$ limit of equation (3.8) is equation (3.7). The circular velocity $v_c(R)$ in the equatorial plane equals $V_0 \bar{R}^{1+\alpha/2}$, so that the value of $V_0$ sets the velocity scale. When $-3 < \alpha < -2$, the potential is $\infty$ at the centre and approaches $\Psi_c$ at large distances. When $\alpha > -2$, the potential equals $\Psi_c$ at the centre and diverges towards $-\infty$ at large distances.

Transformation to the coordinates $(\bar{r},\theta)$ shows that the potentials of the scale–free spheroids, like their densities, are all of separable form

$$\Psi = -V_0^2 \begin{cases} \frac{1}{2}[\ln \bar{r}^2 + P_{-2,q}(\theta)], & \alpha = -2, \\ \frac{1}{\alpha+2}\{\bar{r}^{\alpha+2}\exp[(\alpha+2)P_{\alpha,q}(\theta)] - 1\}, & \alpha \neq -2, \end{cases} \quad (3.11)$$

where $P_{\alpha,q}(\theta)$ is a function of $\theta$ only. It is given explicitly in Appendix B. The case $\alpha = -2$ is a specific example of the general set of scale–free potentials considered by Toomre (1982). We will work with a scaled potential $\overline{\Psi}$ defined as

$$\overline{\Psi} = \begin{cases} 2\Psi/V_0^2, & \alpha = -2, \\ 1 - \Psi/\Psi_c \geq 0, & \alpha \neq -2, \end{cases} \quad (3.12)$$

where $\Psi_c$ is given in equation (3.9).

The density (3.6) and its potential can be combined to express $\tilde{\rho}(\Psi, R^2)$ in terms of a function $\bar{\rho}$ of a single *angular-dependent* variable $\zeta$. We choose to define this variable as

$$\zeta = \begin{cases} \bar{R}^2 \exp(\overline{\Psi}), & \alpha = -2, \\ \bar{R}^2 \overline{\Psi}^{\frac{-2}{\alpha+2}}, & \alpha \neq -2. \end{cases} \quad (3.13)$$

It ranges from 0 on the symmetry axis to 1 in the equatorial plane, where $P_{\alpha,q}(\pi/2) = 0$. We use the definition (3.13) to eliminate $\bar{R}$ dependence from $\tilde{\rho}$, and to bring it to the form

$$\tilde{\rho}(\Psi, R^2) = \frac{\rho_0}{q^\alpha} \bar{\rho}(\zeta) \times \begin{cases} \exp(\overline{\Psi}), & \alpha = -2, \\ \overline{\Psi}^{\frac{\alpha}{\alpha+2}}, & \alpha \neq -2. \end{cases} \quad (3.14)$$

The definition of the function $\bar{\rho}(\zeta)$ which is introduced here is implicit, and is given in equation (B4) of Appendix B. We describe there how to evaluate it numerically for complex $\zeta$.

In order to use the contour integral solution (2.2) we need the value $\xi = \Psi_{\text{env}}(E)$ where the contour crosses the real $\xi$-axis. Use of equations (2.1) gives

$$\Psi_{\text{env}}(E) = \frac{2}{(\alpha+4)}\left(E + \frac{V_0^2}{2}\right). \quad (3.15)$$

It is useful to define two more dimensionless variables, namely the scaled energy $\bar{E}$ and angular momentum $\bar{L}_z$, as

$$\bar{L}_z = \frac{L_z}{bV_0}, \qquad \bar{E} = \begin{cases} 2E/V_0^2 + 1, & \alpha = -2, \\ \frac{2}{(\alpha+4)}(1 - E/\Psi_c), & \alpha \neq -2. \end{cases} \quad (3.16)$$

Substitution in the parametric equations (2.1) for the locus $\mathcal{E}$ of the circular orbits in the $(E, L_z^2)$-plane (Figure 1) then shows that the scaled angular momentum $\bar{L}_c$ of a circular orbit with scaled energy $\bar{E}$ is given by

$$\bar{L}_c^2(\bar{E}) = \begin{cases} \exp(-\bar{E}), & \alpha = -2, \\ \bar{E}^{(\alpha+4)/(\alpha+2)}, & \alpha \neq -2, \end{cases} \quad (3.17)$$

so that the locus $\mathcal{E}$ can be found explicitly in this case.

The DFs of the scale-free spheroids are also of special form. The contour integral solution (2.2) shows that they can be written as

$$f_e(E, L_z) = \frac{\rho_0}{q^\alpha V_0^3} \bar{f}_e(\eta^2) \times \begin{cases} \exp(\bar{E}), & \alpha = -2, \\ \bar{E}^{\frac{\alpha}{(\alpha+2)} - \frac{3}{2}}, & \alpha \neq -2, \end{cases} \quad (3.18)$$

where

$$\eta = \frac{\bar{L}_z}{\bar{L}_c(\bar{E})} = \frac{L_z}{L_c(E)}, \quad (3.19)$$

so that $-1 \leq \eta \leq 1$. Formulas for calculating values of $\bar{f}_e(\eta^2)$ for the different regimes of $\alpha$ are given in Appendix B, together with the elementary expressions for $\bar{f}_e(0)$. These formulas are simplified by our choice (3.13) of the variable $\zeta$, which is closely related to the variable $\eta^2$ in inversion methods (cf. Fricke 1952).

The velocity moments of the scale-free spheroids can be calculated by solution of the Jeans equations, and of the higher-order moment equations. We show in Appendix C that the second moments $\langle v_\phi^2 \rangle$ and $\langle v_R^2 \rangle = \langle v_z^2 \rangle$ are connected by a simple relation (eq. [C2]), and that they can be expressed in terms of elementary functions when $\alpha = -2$.

Equation (3.18) demonstrates that the two-integral DF of the scale-free spheroids is a product of a power of energy and a function that describes the same relative dependence on angular momentum at each energy. This simple form is caused by the scale-free nature of the models. The structure and dynamics at one radius (energy) are related to those at any other radius (energy) by a simple scaling.

The value of $\eta$ indicates the nature of the stellar orbits, from $\eta = 0$ (all orbits with zero angular momentum, which are confined to a meridional plane) to $\eta = \pm 1$ (the circular orbits of maximum angular momentum in the equatorial plane). It has been christened the *circularity* by Gerhard (1991). Figures 4 and 5 show the ratio $\bar{f}_e/\bar{f}_e(0)$ as a function of $\eta^2$ for different values of $\alpha$ and $q$. For oblate spheroids $\bar{f}_e/\bar{f}_e(0)$ is an increasing function of $\eta^2$, while for prolate spheroids it is a decreasing function of $\eta^2$. The range $[\bar{f}_e(0), \bar{f}_e(1)]$ indicates to what extent the high-$L_z$ orbits are needed to maintain the density distribution of the models. At fixed $q < 1$ this range increases as the density profile becomes steeper ($\alpha$ decreases; Figure 4). At fixed $\alpha$ this range increases when the flattening of the oblate scale-free spheroids increases ($q$ decreases; Figure 5), in agreement with results of Dehnen & Gerhard (1994). Flatter models require more nearly circular orbits for self-consistent support. The importance of the high-$L_z$ orbits decreases in prolate models: $\bar{f}_e(1)$ drops below zero for sufficiently large $q$, so that prolate two-integral models are physical only for a limited range of axis ratios (see below).

**Figure 4.** The ratio $\bar{f}_e/\bar{f}_e(0)$ as a function of $\eta^2 = \bar{L}_z^2/\bar{L}_c^2(\bar{E})$ for scale-free spheroids with axis ratio $q = 0.7$ and $\alpha = -2.5, -2, -1.5,$ and $-1$.

**Figure 5.** Same as Figure 4 but for $\alpha = -2$, and $q = 0.4, 0.6, 0.8, 1.0$ and $1.2$. The scale along the vertical axis is logarithmic.

Evans (1993, 1994) constructed a family of self-consistent axisymmetric models with spheroidal potentials rather than spheroidal densities. He took $\Psi \propto (R_E^2 + m^2)^{-\beta_E/2}$ with $m^2 = R^2 + z^2/q_E^2$ and $R_E$ and $q_E$ constants. The $\beta_E = 0$ model has $\Psi = -\frac{1}{2}V_0^2 \ln(R_E^2 + m^2)$. The density of all these models is of the form $\tilde{\rho}(\Psi, R^2) = \rho_0(\Psi) + R^2 \rho_2(\Psi)$, and leads to DFs of the form $f_e(E, L_z) = F_0(E) + L_z^2 F_2(E)$, where $\rho_0, \rho_2, F_0,$ and $F_2$ are elementary functions (powers and exponentials). In the limit of zero core radius, $R_E = 0$, these *power–law models* are scale–free. The density distributions are not spheroidal, become increasingly peanut-shaped when $q_E$ decreases, and are negative along the $z$-axis when $q_E^2 < \frac{1}{2}(1+\beta_E)$. The corresponding DFs are similar to expressions (3.18), with $\alpha = -2-\beta_E$, but have $\tilde{f}_e(\eta^2) = \tilde{f}_e(0)[1+A\eta^2]$, where $A = A(q_E, \beta_E)$. Evans' scale-free power-law models hence are described by straight lines in Figure 4. High–$L_z$ orbits are relatively more important in our spheroidal scale–free models than in Evans' models. This illustrates that in the scale–free models the energy dependence of $f_e$ is determined completely by the slope $\alpha$ of the density profile. The $\eta$–dependence, on the other hand, is influenced by $\alpha$, by the flattening $q$ (or $q_E$), and by the shape of the surfaces of constant density.

The density $\tilde{\rho}(\Psi, R^2)$ of the scale–free spheroids can be expanded in powers of $R^2$ times functions of $\Psi$ alone. Equivalently, the function $\bar{\rho}(\zeta)$ can be expanded in powers of $\zeta$. Unlike Evans' power-law models, this expansion has more than two terms, and, because of the direct relationship between powers of $\zeta$ in $\bar{\rho}(\zeta)$ and powers of $\eta^2$ in $\tilde{f}_e$ (cf. Appendix B, eqs. [B8] through [B11]), the corresponding series in $\eta^2$ for $\tilde{f}_e$ also has more than two terms. Calculation of the successive terms becomes rapidly unwieldy, so that evaluation of $\tilde{f}_e$ by means of the HQ method, as we have done, is more practical. Figure 5 shows that, to first order, the scale–free spheroids have $\tilde{f}_e(\eta) \sim \tilde{f}_e(0)\exp(A\eta^2)$, with $A = A(q, \alpha)$ a constant. This approximation is quite accurate for $q$ in the range between 0.75 and 1.2. It suggests that scale–free models with an *exact* exponential dependence on $\eta^2$ have nearly spheroidal densities.

Figure 6 shows the region in the $(q, \alpha)$-plane where the even two–integral DF (3.18) of the scale–free spheroids is non–negative. All oblate spheroids of this kind have $f_e(E, L_z) \geq 0$. However, at fixed $\alpha$ there is a maximum axis ratio $q_{\max}(\alpha) > 1$ beyond which $f_e(E, L_z) < 0$ for prolate models. This is in harmony with earlier studies of specific two–integral axisymmetric models (e.g., Dejonghe & de Zeeuw 1988; Batsleer & Dejonghe 1993), and is caused by the fact that the $\eta = 0$ orbits needed to reproduce the density along the long axis of the model overpopulate the density in the equatorial plane when the model becomes sufficiently elongated. The derived DF corrects this overpopulation by giving the $\eta = 1$ circular orbits negative weight, and hence is unphysical. The range of physical scale–free $f_e(E, L_z)$ prolate models decreases when $\alpha$ decreases, i.e., when the density profile steepens. At fixed $q$ the potential becomes more nearly spherical when $\alpha$ decreases, and so do the orbital densities, so that the danger of overpopulation of the equatorial plane increases. Similar results were found by Evans (1994) for the scale–free power–law models. His Figure 1 shows a $(q_E, \beta_E)$-diagram ($\beta_E = -2 - \alpha$) which can be compared to our Figure 6 (but note that $q_E$ is the axis ratio of the potential, and not of the density). Physical prolate power-law models have a maximum allowed axis ratio. However, oblate power-law models have $f_e(E, L_z) \geq 0$ only when $q_E^2 \geq \frac{1}{2}(1+\beta_E)$.

### 3.3 Small radii: spheroidal cusps and black holes

For $\alpha < 0$, the density (3.1) of the $(\alpha, \beta)$-models has a power-law cusp near the centre, $\rho \approx \rho_0(m/b)^\alpha$, approximating the density of the scale–free spheroids. Since the presence of a central black hole affects significantly the behaviour of the potential near the centre, we discuss cases with and without it separately.

**Figure 6.** The horizontally hatched region indicates the area in the $(q, \alpha)$ plane where the scale–free spheroids have physical DFs $f_e(E, L_z)$. All oblate ($q < 1$) models have $f_e(E, L_z) \geq 0$. At fixed $\alpha$, physical two–integral prolate models do not exist beyond a maximum elongation. The diagonally hatched area indicates the more limited region where such spheroids have a physical $f_e(E, L_z)$ in the presence of a central black hole.

When there is no black hole, the potential can be approximated by that of a scale–free spheroid, provided that the contribution to the potential from the power–law cusp dominates contributions by the rest of the system. This occurs for systems with parameters $(\alpha, \beta)$ such that $\alpha \leq -2$, or $\alpha > -2$ and $\alpha + 2\beta \geq -2$, i.e., outside the horizontally hatched region in Figure 3. To within an additive constant, the potential is given by (3.7) for systems with $\alpha = -2$, and by (3.8) otherwise. For these systems the entire analysis of the scale–free spheroids is applicable once the potential $\Psi$ and the energy $E$ are both modified by the additive constant, so the DF for high energy stars near the centre can be readily calculated with the results of Section 3.2 and Appendix B.

Stars outside the central region contribute significantly to the finite central potential (3.4) when $\alpha > -2$ and $\alpha + 2\beta < -2$. Then, the correct approximation to the potential near the centre is, using the second formula in equation (2.8), $\Psi = \Psi_0^* - \Psi_c + \Psi^{\text{cusp}}$, where $\Psi^{\text{cusp}}$ and $\Psi_c$ are given in equations (3.8) and (3.9), respectively. We now observe that $\rho = \rho_0(m/b)^\alpha$ and $\Psi - \Psi_0^* + \Psi_c$ form a pair of scale-free spheroids with power index $\alpha$. Hence we only need to replace $\Psi$ and $E$ in the relevant formulas of the previous section, by $\Psi - \Psi_0^* + \Psi_c$ and $E - \Psi_0^* + \Psi_c$, respectively. For example, the dimensionless energy $\bar{E}$ in equations (3.16) becomes

$$\bar{E} = \frac{2(\alpha+2)}{(\alpha+4)}\frac{(\Psi_0^* - E)}{V_0^2}. \tag{3.20}$$

When a central black hole is present, its potential overwhelms the stellar potential at sufficiently small radii. An asymptotic expression for $f_e(E, L_z)$ can be calculated as the DF needed to maintain the power-law cusp density in the central black hole potential. In most cases the result of including the stellar potential, even in its approximate form, is that we are no longer able to obtain a simple, explicit, function $\tilde{\rho}(\Psi, R^2)$. There is one exception, which is again provided by systems with $\alpha > -2$ and $\alpha + 2\beta < -2$. Then the central stellar potential is finite and can be added to the black hole potential to approximate the potential at small radii by

$$\Psi(R^2, z^2) = \frac{GM_{\rm BH}}{\sqrt{R^2 + z^2}} + \Psi_0^*, \quad (3.21)$$

where $\Psi_0^*$ is the finite central potential (3.4) for the special cases mentioned above, and zero otherwise. The first term always dominates at sufficiently small radii, but retaining the second term provides a more accurate approximation with no extra work. Upon solving $z^2$ from equation (3.21) and substituting the result into the scale-free cusp density, we obtain an approximate, yet explicit, relation

$$\tilde{\rho}(\Psi, \bar{R}^2) = \frac{\rho_0}{q^\alpha} \left(\frac{\Psi - \Psi_0^*}{B}\right)^{-\alpha} \left[1 - e^2 \bar{R}^2 \left(\frac{\Psi - \Psi_0^*}{B}\right)^2\right]^{\alpha/2}, \quad (3.22)$$

with $B = GM_{\rm BH}/b$ a reference potential and $e^2 = 1 - q^2$. This density $\tilde{\rho}(\Psi, \bar{R}^2)$ is a minor generalization of a component introduced by Dejonghe (1986). HQ give a real one-dimensional integral formula for its DF, their eq. (B18), from which a factor of 1/2 is missing (Dehnen & Gerhard 1994). The DF for the density (3.22) is readily obtained by replacing $E$ by $E - \Psi_0^*$ in that formula to give

$$f_e(E, L_z) = \frac{\rho_0 (B/q)^\alpha}{4\pi^2} \frac{\partial^2}{\partial E^2} \left\{ (2E - 2\Psi_0^*)^{1/2-\alpha} \times \int_0^\pi \cos\left[\tfrac{1}{2}(1-\alpha)\theta\right] \left[\frac{1}{\cos^2\theta/2} - e^2\eta^2\right]^{\alpha/2} d\theta \right\} \quad (3.23)$$

$$= \frac{\rho_0 q^{-\alpha}}{B^{3/2}} \left(\frac{E - \Psi_0^*}{B}\right)^{-\alpha - 3/2} \bar{f}_\alpha(e^2\eta^2),$$

where the function $\bar{f}_\alpha$ is an elementary real integral, which can be written in terms of a generalized hypergeometric function (Dejonghe 1986)

$$\bar{f}_\alpha(e^2\eta^2) = \frac{1}{(2\pi)^{3/2}} \frac{\Gamma(1-\alpha)}{\Gamma(-\alpha - 1/2)} \quad (3.24)$$
$$\times \,_3F_2(\tfrac{1-\alpha}{2}, 1 - \tfrac{\alpha}{2}, -\tfrac{\alpha}{2}; -\alpha - \tfrac{1}{2}, \tfrac{1}{2}; e^2\eta^2).$$

The DF (3.23) is again of a separable form: the product of a power of $(E - \Psi_0^*)/B$ and the function $\bar{f}_\alpha$ whose argument is $e^2\eta^2 = e^2 L_z^2/L_c^2(E)$, where $L_c^2(E) = (GM_{\rm BH})^2/[2(E - \Psi_0^*)]$ is the square of the angular momentum of a circular orbit of energy $E$. The function $\bar{f}_\alpha$ depends only on the power $\alpha$ of the density cusp, and not on the axis ratio $q$, though this ratio is part of the argument of $\bar{f}_\alpha$. For integer $\alpha$ the generalized hypergeometric function in the definition of $\bar{f}_\alpha$ reduces to an ordinary hypergeometric function, which can be written in terms of elementary functions. Table 1 lists $\bar{f}_\alpha$ for the cases $\alpha = -1, -2$ and $-3$, which have very simple forms. The case $\alpha = -4$ is given in Appendix B of Dehnen & Gerhard (1994). For integer $\alpha < -4$ the expressions rapidly become lengthy.

The value of $\bar{f}_\alpha(0)$ is the elementary factor in front of $_3F_2$ in equation (3.24), so that $f_e(E, 0)$ is elementary. For spherical cusps ($q = 1$, $e = 0$), $f_e(E, 0)$ gives the isotropic $f(E)$ for large energy $E$. It agrees with the expression given by Tremaine et al. (1994) for Dehnen's (1993) family of cusped spherical models (but they neglect the term $\Psi_0^*$). For oblate cusps $\bar{f}_\alpha$ is a monotonically increasing function of its argument, and the DF (3.23) is least for stars with zero angular momentum ($\eta = 0$). The extreme value $\bar{f}_\alpha(e^2)$ becomes large as $e$ grows and the isodensity contours flatten. For $f_e(E, 0)$ to be non-negative, we must have $\alpha \leq -1/2$. This is the same constraint on the slope of the cusp density profile as in the spherical case (Tremaine et al. 1994). In prolate cusps $e^2 = 1 - q^2 < 0$, and the DF (3.23) is least for stars on circular orbits ($\eta = 1$). It is, of course, unphysical for $\alpha > -1/2$, but for smaller $\alpha$ it is physical only when $q$ is less than a certain maximum value. The region in the $(q, \alpha)$-plane where spheroidal cusps have physical $f_e(E, L_z)$ in the presence of a central black hole is shown in Figure 6. For prolate models this allowed region is more limited than the area occupied by physical self-consistent $f(E, L_z)$ scale-free spheroids. Near the black hole the stars experience a spherical potential. As we have seen in Section 3.2, this increases the danger of overpopulating the density in the equatorial plane of a prolate model, so that at fixed $\alpha$ the physical range of $q > 1$ is smaller. The potential of the self-consistent spheroids becomes more nearly spherical (and Keplerian) when $\alpha$ decreases towards $-3$ at fixed $q$, so that the shrinking of the allowed range of $q$ caused by the inclusion of the black hole is less.

**Table 1.** Some special cases of $\bar{f}_\alpha$.

| $\alpha$ | $\bar{f}_\alpha(e^2\eta^2)$ |
|---|---|
| $-1$ | $\frac{1}{2\sqrt{2}\pi^2} \frac{1 + e^2\eta^2}{(1 - e^2\eta^2)^2}$ |
| $-2$ | $\frac{1}{\sqrt{2}\pi^2} \frac{(2 + e^2\eta^2)\sqrt{1 - e^2\eta^2} + 3e\eta \arcsin e\eta}{(1 - e^2\eta^2)^{5/2}}$ |
| $-3$ | $\frac{2\sqrt{2}}{\pi^2} \frac{1 + 3e^2\eta^2}{(1 - e^2\eta^2)^3}$ |

### 3.4 Large radii: power-law halos

At large radii the density approaches $\rho = \rho_0 (m/b)^{\alpha + 2\beta}$, which is again scale-free. We need to distinguish systems of finite mass ($\alpha + 2\beta < -3$; the vertically hatched area in Figure 3) and those of infinite mass ($\alpha + 2\beta > -3$). We ignore the intermediate $\alpha + 2\beta = -3$ case in which the total mass is marginally infinite. For the former, both the stellar potential and that of any central black hole are Keplerian and both should be included in an approximate potential, unless $M_{\rm BH} \ll M$. The DF is again given by formula (3.23) with $\Psi_0^* = 0$, $\alpha$ replaced by $\alpha + 2\beta$, and the total stellar mass $M$ (3.5) added to $M_{\rm BH}$ in the definition of $B$.

In a self-consistent system of infinite mass, the stellar potential at large radii dominates any Keplerian potential, and so is insensitive to any central black hole. Therefore the potential approximation for cases with or without a central black hole is the same. Hence, all systems with $\alpha + 2\beta > -3$ are approximated at large radii by scale-free spheroids of the type discussed in section 3.2; the approximate DF $f_e(E, L_z)$ for stars at large radii ($E \to \Psi_\infty$) can be obtained once we replace $\alpha$ by $\alpha + 2\beta$ and modify both $\Psi$ and $E$ by additive constants if necessary.

We have also investigated the case of a flattened spheroidal density $\rho = \rho_0 (m/b)^\alpha$, with $m^2 = R^2 + z^2/q^2$, in the non-self-consistent scale–free power–law potential $\Psi_d$ defined as (cf. Evans 1994)

$$\Psi_d = -V_0^2 \begin{cases} \ln \overline{m}_d, & \gamma = 0, \\ \frac{1}{\gamma}[\overline{m}_d^\gamma - 1], & \gamma \neq 0, \end{cases} \quad (3.25)$$

where $\overline{m}_d = m_d/c$, with $m_d^2 = R^2 + z^2/q_d^2$, so that $q_d$ is the axis ratio of the spheroidal equipotentials and $c$ is a reference length. When $q_d = 1$ and $\gamma = -1$ this reduces to the case discussed in Section 3.3 of a spheroidal density $\rho_0(m/b)^\alpha$ in the Kepler potential. Proceeding as in that case, we again find an explicit expression for $\tilde{\rho}(\Psi_d, \bar{R}^2)$ which is another generalization of the Dejonghe (1986) component (cf. eq. [3.22]). When $\gamma \neq 0$, it is given by

$$\tilde{\rho}(\Psi_d, \bar{R}^2) = \rho_0 \left(\frac{cq_d}{bq}\right)^\alpha \left(1 - \frac{\gamma \Psi_d}{V_0^2}\right)^{\alpha/\gamma} \times \\ \left[1 - e_d^2 \bar{R}_d^2 \left(1 - \frac{\gamma \Psi_d}{V_0^2}\right)^{-2/\gamma}\right]^{\alpha/2}, \quad (3.26)$$

where $e_d^2 = 1 - q^2/q_d^2$ and $\bar{R}_d = R/c$. When $\gamma = 0$, so that the potential is logarithmic, we find

$$\tilde{\rho}(\Psi_d, \bar{R}^2) = \rho_0 \left(\frac{cq_d}{bq}\right)^\alpha \exp(-\alpha \Psi_d/V_0^2) \times \\ \left[1 - e_d^2 \bar{R}_d^2 \exp(2\Psi_d/V_0^2)\right]^{\alpha/2}. \quad (3.27)$$

Application of the HQ method shows that the corresponding $f_e$ is given by

$$f_e(E, L_z) = \frac{\rho_0}{V_0^3} \left(\frac{cq_d}{bq}\right)^\alpha \tilde{f}_{\alpha,\gamma}(e_d^2 \eta^2) \times \begin{cases} \exp(-\frac{\alpha \bar{E}}{2}), & \gamma = 0, \\ \bar{E}^{\frac{\alpha}{\gamma} - \frac{3}{2}}, & \gamma \neq 0, \end{cases} \quad (3.28)$$

where $\bar{E}$ is defined as

$$\bar{E} = \begin{cases} \frac{2E}{V_0^2} + 1, & \gamma = 0, \\ \frac{2}{\gamma + 2}(1 - \frac{\gamma E}{V_0^2}), & \gamma \neq 0. \end{cases} \quad (3.29)$$

The $\tilde{f}_{\alpha,\gamma}(e_d^2 \eta^2)$ can be evaluated in the same way as the function $\tilde{f}_e$ of equation (3.18), as detailed in Appendix B. These DFs are useful for understanding the behaviour of two–integral models in the limit of large radii, where the density falls off as a power–law, and the potential may be dominated by a dark halo. We remark that changing the axis ratios $q$ and $q_d$ as well as the normalisation lengths $b$ and $c$ while keeping both $q/q_d$ and $b/c$ constant leaves the DF invariant.

## 4 APPLICATION TO M32

To illustrate our technique we have used it to model the galaxy M32, which is believed to harbour a massive black hole in its nucleus (Tonry 1987; Dressler & Richstone 1988; Richstone, Bower & Dressler 1990). HST observations have revealed the presence of a central surface brightness cusp (Lauer et al. 1992). High spatial resolution (ground–based) kinematical and VP measurements along several position angles are available from van der Marel et al. (1994a, hereafter vdM94a). Axisymmetric $f(E, L_z)$ models were used by van der Marel et al. (1994b, hereafter vdM94b) to interpret these observations. The modelling consisted of: (i) use of Evans' (1994) power–law model DFs without a central black hole; and (ii) calculation of the first three moments of the VP for an $(\alpha, \beta)$–model with a black hole, by solution of the moment equations of the collisionless Boltzmann equation. A remarkably good fit was obtained with a black hole of mass $M_{\rm BH} \approx 1.8 \times 10^6 \, {\rm M}_\odot$, but the actual DF of the model could not be calculated. With the technique presented here we *can* calculate the entire DF (Section 4.1). This in turn yields the full VP shapes, and hence allows a more accurate comparison to the data of vdM94a (Section 4.2). It also allows a detailed study of the expected VP shapes for the high spatial resolution spectroscopic observations that are soon to be expected with the HST (Section 4.3).

**Figure 7.** Contour plot of the even part $f_e(E, L_z)$ of the DF of our model for M32, which has a central black hole of mass $1.8 \times 10^6 \, {\rm M}_\odot$, as discussed in the text. The quantities along the axes are $\eta^2 = L_z^2/L_c^2(E)$, and the energy $E$ in units of $\Psi_0^*$, which is the central potential in the absence of a black hole. The allowed energy range is $E \in [0, \infty)$, but only the range $[0.1, 2]$ is shown. At lower and higher energies the DF has reached its asymptotic behaviour as dictated by the scale–free approximations. Adjacent solid contours are a factor 0.17 apart, the highest 'contour' being at the maximum of the DF, which is indicated by the solid dot. The ten dashed contours are a factor $(0.17)^{0.1}$ apart, the highest contour again being at the maximum of the DF.

### 4.1 The $f(E, L_z)$ distribution function for M32

The observed surface brightness distribution of M32 can be well fitted with an $(\alpha, \beta)$–model with parameters $\alpha = -1.435$, $\beta = -0.423$, $b = 0.55''$, $\rho_0 = j_0 \Upsilon_V \, {\rm M}_\odot / {\rm L}_{\odot, V}$, $j_0 = 0.470 \times 10^5 (q'/q) \, {\rm L}_{\odot, V} \, {\rm pc}^{-3}$ and $q' = 0.73$. This leaves three free parameters that can be chosen to optimize the fit to the kinematical observations: the inclination $i$ (which enters in the relation [3.3] between $q$ and $q'$), the average $V$-band mass–to–light ratio $\Upsilon_V$ of the stellar population in solar units (assumed to be independent of radius), and the black hole mass $M_{\rm BH}$. We restrict ourselves here to the model that is edge-on ($i = 90°$) and has $\Upsilon_V = 2.51$ and $M_{\rm BH} = 1.8 \times 10^6 \, {\rm M}_\odot$, based on the results of vdM94b. We do not discuss the detailed dependence of the model predictions on the parameters $i$, $\Upsilon_V$ and $M_{\rm BH}$, since we do not expect the results of our models to change the discussion given previously by vdM94b, on the basis of their more approximate modelling.

Figure 7 shows a contour diagram of the even part $f_e(E, L_z)$ of the DF for our model for M32, obtained with the technique described in Sections 2 and 3. The DF is positive for all physical values of $(E, L_z)$. The contours slope gently upwards in the right half in accordance with the weak dependence of $f_e$ on energy predicted by the scale–free approximation (3.23) for our model with $\alpha = -1.435$ (which is close to $-1.5$, the value for which the energy dependence in eq. [3.23] vanishes). The contours slope sharply downwards in the left half in accordance with the strong dependence of $f_e$ on energy predicted by the scale–free approximation (3.18) for the appropriate $\alpha$ value of $-2.281$. Figure 8, which shows the dependence of $f_e$ on the energy $E$, for both $L_z = 0$ and $L_z = L_c(E)$, confirms the accuracy of the scale–free approximations for the limit of low and high energies. Figure 9 shows a contour plot of $f_e(E, L_z)/f_e(E, 0)$. The nearly horizontal contours indicate that the $f_e$ for our M32 model is close to being a separable function of $E$ and $\eta^2 = L_z^2/L_c^2(E)$ at all energies, with relatively small discrepancies in the transition region $0.5 \lesssim E \lesssim 1.1$ between low and high energies. The similarity of the behaviour of $f_e(E, L_z)/f_e(E, 0)$ as a function of $\eta^2$ in the low and the high energy limit is further illustrated in Figure 10, which shows the asymptotic behaviour obtained from the scale–free approximations. For comparison, the predicted behaviour at high energies for the same model without the central black hole is also shown. The dependence of $f_e(E, L_z)/f_e(E, 0)$ as function of $\eta^2$ is much steeper at high energies when the black hole is present, because then the stars close to the centre orbit in a spherical rather than a flattened potential. This causes the density contributed by stars of the same $(E, L_z)$ to be more nearly round than in the case of a flattened potential (Section 3.4). In order to reproduce the same flattened density distribution, the number of stars on high–$L_z$ orbits must therefore increase relative to the case where the potential is flattened.

Apparently, the mass density slope in the inner parts of M32 and the presence of the black hole 'conspire' to produce nearly the same dependence of $f_e(E, L_z)/f_e(E, 0)$ on $\eta^2$ at high energies, as at low energies, where the dependence is governed by the mass density slope in the outer parts. It is not clear whether this is a mere coincidence, or is the result of stellar dynamical processes which have operated in M32, possibly caused by the presence of the central black hole. Such processes would then have to be capable of removing any dependence of the DF on a third integral of motion *and* would have to drive the DF to a product form, all in less than the Hubble time. More quantitative theoretical work, e.g., by study of the adiabatic growth of central black holes in stellar systems (Young 1980; Quinlan, Hernquist & Sigurdsson 1994), is clearly needed in this area.

### 4.2 Comparison to ground–based kinematical data

To compare the kinematical predictions of our model to the available data we must specify also the odd part $f_o$ of the

**Figure 8.** Behaviour of $f_e(E, L_z)$ for our model for M32 discussed in the text, as function of energy $E$, for $\eta^2 = 0$ ($L_z = 0$; solid line) and for $\eta^2 = 1$ ($L_z = L_c(E)$; dashed line). The dotted lines show the asymptotic slopes at low and high energies expected from the scale–free approximations. These slopes agree well with the model slopes. The actual *values* predicted by the scale–free approximations also agree well with the model predictions (for all $\eta^2$), but are not shown here for the purpose of clarity.

**Figure 9.** Contour plot of $f_e(E, L_z)/f_e(E, 0)$ for our model for M32 discussed in the text. The quantities along the axes are as in Figure 7. Adjacent contours in the figure differ by a factor 0.8. The highest contours lie at the top of the plot.

DF. We choose the parametrization

$$f_o(E, L_z) = (2F - 1)\frac{\tanh[a\eta/2]}{\tanh[a/2]} \, f_e(E, L_z), \qquad (4.1)$$

where as before $\eta = L_z/L_c(E)$, and $0 \leq F \leq 1$ and $a > 0$ are free parameters. This is a modified version of a functional form derived by Dejonghe (1986) based on maximum entropy arguments. With this choice for $f_o$, the total DF

**Figure 10.** The behaviour of $f_e(E, L_z)/f_e(E, 0)$ for our model for M32 discussed in the text, as a function of $\eta^2 = L_z^2/L_c^2(E)$, in the limit of low energies (solid curve) and high energies (dashed curve). The dotted curve shows what the behaviour in the limit of high energies would have been without the central black hole.

$f = f_e + f_o$ is positive whenever $f_e$ is. We adopt $F = 1$ and $a = 5.5$, based on the results of vdM94b for Evans' power–law models. The DF of our model for M32 is now specified completely, and the model VPs can be calculated as described in Section 2.3. Figure 11 compares the model predictions to the data along five different slit positions presented by vdM94a. Each VP is characterized by six parameters: the mean $V$ and dispersion $\sigma$ of the best-fitting Gaussian to the VP, and the Gauss–Hermite moments $h_3, \ldots, h_6$, defined as in van der Marel & Franx (1993). Our model predictions take the spatial binning and seeing point spread function (PSF) convolution of the observations into account, as described in Appendix D. Small ($\lesssim 0.1''$) offsets of the slit from the galaxy centre due to differential atmospheric refraction were also modelled. The results in Figure 11 confirm the main conclusions of the modelling by vdM94b. The model fits the data remarkably well, much better than one would have expected a priori. The observed steep central gradient in the mean velocity and the observed central peak in the velocity dispersion are both reproduced (owing to the presence of the central black hole in the model). The observed Gauss–Hermite coefficients are fitted up to a rms residual of only $\sim 0.02$, indicating that the dynamical structure of M32 is most likely very close to that of an $f(E, L_z)$ model. Nonetheless, some minor discrepancies between the observations and the model predictions remain, most likely indicating a (slight) dependence of the DF on a third integral.

First, the small discrepancies between the observed and the predicted Gauss–Hermite coefficients outside the central arcsec on the major axis, are in the sense that the even part of the observed major axis VPs is slightly more flat-topped than predicted. This might indicate that M32 has a velocity distribution with $\langle v_\phi^2 \rangle > \langle v_\theta^2 \rangle \gtrsim \langle v_r^2 \rangle$. The discrepancy is hardly significant, however, and is in fact smaller than that inferred by vdM94b. This we have found to be due to the fact that the $(\alpha, \beta)$–model used here has an asymptotic mass-density slope that is slightly steeper than that of the power–law model employed by vdM94b ($\rho \propto m^{-2.28}$ versus $\rho \propto m^{-2.2}$). It is not a consequence of the fact that the iso-density contours of the $(\alpha, \beta)$– and power–law models have slightly different shapes.

Secondly, Figure 11 shows that the predicted amplitude of the mean line-of-sight motion $V$ along the intermediate axes is slightly too high. This is consistent with the findings of vdM94b, who argued that to obtain a good fit on both axes, one must invoke an odd part of the DF that depends also on a third integral. To test this conclusion, we attempted to solve the inverse problem. We adopted a streaming velocity field defined by $\langle v_\phi \rangle^2 = k(\langle v_\phi^2 \rangle - \langle v_R^2 \rangle)$ with $k = 1.25$, which provides a good fit to $\langle v_{\rm los} \rangle$ on both the major and the intermediate axes (vdM94b). We then used the HQ method to obtain the unique $f_o(E, L_z)$ consistent with this streaming velocity field. We found that the resulting model is unphysical, because the total DF $f = f_e + f_o$ is not positive for all physically accessible $(E, L_z)$. It thus appears indeed, that an odd part of the form $f_o = f_o(E, L_z, I_3)$ is required to fit the amplitude of the mean streaming motions along all slit positions simultaneously.

Thirdly, even with the inclusion of a central black hole there remains a discrepancy between the observed and the predicted velocity dispersions near the centre. Especially on the minor axis, the observed central peak in the velocity dispersion is steeper than that predicted by the model. This hints at a dependence of the even part of the DF on a third integral.

The most important conclusion from the $f(E, L_z)$ modelling of M32, however, is that, aside from the minor discrepancies discussed above, the accuracy with which an $f(E, L_z)$ model can fit the data is quite remarkable. A similar conclusion was reached independently by Dehnen (1995), who used a Richardson–Lucy algorithm (Newton & Binney 1987) to construct $f(E, L_z)$ models for M32. Our results are not very sensitive to the assumed inclination angle (vdM94), which is not well constrained by the data. It is also not very sensitive to the precise value of the slope of the density profile inside $0.3''$, which might differ slightly from the value adopted here (Lauer et al. 1992). With a slightly steeper slope an equally good fit is obtained with a slightly less massive black hole, and vice-versa.

### 4.3 Predictions for observations with HST

The fact that an $f(E, L_z)$ model with a central black hole can provide such a good fit to ground–based kinematical and VP data does not necessarily imply that M32 must have a central black hole. To date it has not been convincingly demonstrated that three–integral axisymmetric models without a central black hole cannot also fit the same ground–based data. High spatial resolution data from the refurbished HST should provide more definite evidence either for or against the presence of a central black hole. In this section we discuss the kinematical and VP predictions of our model for M32, for observations through the small apertures available on the HST. This yields definite predictions for the signatures of a central black hole that one might expect to observe with the HST.

We discuss the normalization $\gamma$, the mean $V$ and the dispersion $\sigma$ of the best-fitting Gaussian to the VP, as well as the Gauss–Hermite moments up to order 6. In a real observational situation a galaxy spectrum is modelled as the convolution of a (stellar) template spectrum and a broadening function. The ratio of the equivalent width of the absorption lines in the galaxy spectrum to those in the template spectrum is called the 'line strength'. This line strength is unknown, and has to be estimated from the data. If this true line strength is one, then our parameter $\gamma$ is the estimate of the 'line strength' that one would expect to obtain by fitting a Gaussian broadening function to the data.

**Figure 11.** The data points are the observed kinematics and VP parameters for M32 as a function of the projected radius $R'$ along five different slit position angles, as presented by vdM94a. From top to bottom: the mean and dispersion of the best–fitting Gaussian to the VP, and the Gauss–Hermite coefficients $h_3, \ldots, h_6$. The curves show the predictions of our model for M32, which has a $1.8 \times 10^6 \, M_\odot$ central black hole, taking into account the seeing convolution and spatial binning of the observations.

**Figure 12.** Predicted kinematics and VP parameters for our model for M32, which has a $1.8 \times 10^6 \, M_\odot$ central black hole, for observations through a circular aperture placed on the galaxy centre. Solid curves show as functions of the aperture diameter $D$: the normalization $\gamma$ and dispersion $\sigma$ of the best-fitting Gaussian to the VP, and the Gauss–Hermite moments $h_4$ and $h_6$. The short-dashed curves in the left panels show the true normalization and dispersion of the VP.

*4.3.1 Predictions for centred apertures*

We first calculated the predicted VPs for observations through a circular aperture placed on the galaxy centre. Figure 12 shows the predicted $(\gamma, \sigma, h_4, h_6)$ as functions of the aperture diameter $D$ [the quantities $(V, h_3, h_5)$ are zero since the central VPs are symmetric]. The true line strength (assumed to be $\equiv 1$) and dispersion of the VP are shown also.

The true velocity dispersion of the VP increases with decreasing $D$, roughly as $\sigma^2 \approx c_1 + c_2/D$, where $c_1$ and $c_2$ are constants. For $D \gtrsim 0.5''$, the predicted VP is close to Gaussian. For smaller diameters the VP wings are more extended than those of a Gaussian (see also Figure 15 below). This is due to the stars that orbit close to the hole at very high velocities, and is quantified by the increasingly non-zero values of $h_4$ and $h_6$. The non-Gaussian wings of the VP contribute significantly to the normalization and dispersion of the VP. A Gaussian fit is insensitive to the wings of the VP, and hence underestimates both the true line strength and the true velocity dispersion.

**Figure 13.** Predicted kinematics and VP parameters for our model for M32, which has a $1.8 \times 10^6 \, M_\odot$ central black hole, for observations through a $0.09'' \times 0.09''$ square aperture, placed along the major axis at a distance $R'$ from the galaxy centre. Solid curves show as functions of $R'$: the normalization $\gamma$, mean $V$ and dispersion $\sigma$ of the best-fitting Gaussian to the VP, and the Gauss–Hermite moments $h_3, \ldots, h_6$. The short-dashed curves in the left panels show the true normalization, mean and dispersion of the VP. The ground-based major axis $V$ and $\sigma$ observed by vdM94a are also shown for comparison (dots). The long-dashed curve shows the model fit to these data. The results illustrate the major improvement to be expected with the HST.

Our predictions for M32 are qualitatively similar to those of van der Marel (1994d), who discussed the expected kinematics and VP shapes for centred aperture observations of the galaxy M87, using a spherical model with a $5 \times 10^9 \, M_\odot$ central black hole. Nonetheless, there are a few noticeable differences. The stars that are influenced significantly by a

**Figure 14.** As Figure 13, but now for observations through a circular aperture of diameter $D = 0.26''$, placed along the major axis at a distance $R'$ from the galaxy centre.

central black hole in a stellar system reside in a sphere with radius of order $r_{\rm BH} \equiv 2GM_{\rm BH}/3\pi\overline{\sigma}^2$, where $\overline{\sigma}$ is a 'typical' velocity dispersion outside the region influenced by the black hole. With this definition, the projected velocity dispersion $\sigma$ of a singular isothermal sphere with a massive black hole satisfies: $\sigma^2 = \overline{\sigma}^2[1 + (r_{\rm BH}/R')]$ (Tremaine et al. 1994). In arcseconds on the sky:

$$r_{\rm BH} = 0.019'' \left(\frac{M_{\rm BH}}{10^6 M_\odot}\right)\left(\frac{100\,{\rm km\,s^{-1}}}{\overline{\sigma}}\right)^2\left(\frac{1\,{\rm Mpc}}{d}\right), \qquad (4.2)$$

where $d$ is the distance to the galaxy. This radius is approximately four times smaller for M32 ($M_{\rm BH} \approx 1.8 \times 10^6 M_\odot$, $\overline{\sigma} \approx 55\,{\rm km\,s^{-1}}$, $d \approx 0.7\,{\rm Mpc}$) than for M87 ($M_{\rm BH} \approx 5 \times 10^9 M_\odot$, $\overline{\sigma} \approx 300\,{\rm km\,s^{-1}}$, $d \approx 16\,{\rm Mpc}$). So to detect similar deviations from a Gaussian for both galaxies, M32 must be observed with a four times smaller aperture than M87. Conversely, for a fixed aperture size, the expected VP for M32

**Figure 15.** The two solid curves are the VPs predicted by our model for M32, which has a $1.8 \times 10^6 M_\odot$ central black hole, for observations through a $0.09'' \times 0.09''$ square aperture: (i) placed on the galaxy centre ($R' = 0''$); and (ii) placed along the major axis at $R' = 0.1''$. Both VPs are normalized. The two dotted curves are the best-fitting Gaussians to these VPs. The arrows indicate the central escape velocity $\pm\sqrt{2\Psi_0^*}$, due to the gravitational potential of the stars. If there were no black hole in the model, no stars would be observed beyond this velocity.

is much closer to a Gaussian than that for M87.

From an observational point of view, there are two main differences between M32 and M87. First, M32 has a much smaller velocity dispersion. So while for M87 the continuum subtraction in the spectral analysis is a serious problem (van der Marel 1994c,d), this is not expected to be the case for M32. On the other hand, for M32 the limited instrumental resolution will be a complicating factor. The Faint Object Spectrograph (FOS) aboard the HST has $\sigma_{\rm instr} \approx 100\,{\rm km\,s^{-1}}$, of the same order as the stellar velocity dispersion. The second important difference between M32 and M87 is that M32 has a much higher surface brightness. For observations of the M87 centre with a $D = 0.26''$ aperture, exposure times of $\gtrsim 10$ hours are required to obtain a sufficient signal–to–noise ratio for a useful VP analysis. For M32, not more than $\sim 15$ minutes are required.

*4.3.2 Predictions for apertures placed along the major axis*

To obtain constraints on the rotational properties of M32 it will be useful to obtain HST aperture observations at various distances along the major axis. We therefore calculated the predicted VPs of our model for M32, for observations with: (i) a $0.09'' \times 0.09''$ square aperture; and (ii) a $D = 0.26''$ circular aperture. The first is the size of the smallest aperture available on the HST/FOS, the second is the size of the smallest circular aperture available on the HST/FOS. Figures 13 and 14 show the predicted $(\gamma, V, \sigma, h_3, \ldots, h_6)$ as functions of the galactocentric distance $R'$ on the sky. As in Figure 12, the true normalization, mean and dispersion of the VP are shown for comparison, as well as the observed and predicted ground-based $V$ and $\sigma$ (from Figure 11). Figure 15 shows the predicted VPs with the square aperture for $R' = 0''$ and $R' = 0.1''$, together with the best Gaussian fits to these VPs.

The velocity dispersion one would expect to measure in the centre ($R' = 0''$) by fitting a Gaussian VP to the data, is 127 km s$^{-1}$ for the square aperture and 105 km s$^{-1}$ for the circular aperture. This is significantly larger than the central velocity dispersion of $\sim 86$ km s$^{-1}$ obtained from ground-based data (see Figure 11). Figure 15 shows that the broad wings of the central VP provide a strong signature of the black hole. The arrows in the figure indicate the central escape velocity $\sqrt{2\Psi_0^*}$ due to the gravitational potential of the stars, which for our model is 298 km s$^{-1}$. In the absence of the central black hole no stars would be observed beyond this velocity.

Outside the centre ($R' \neq 0$) the predicted VPs are asymmetric, with a tail away from the direction of rotation ($V$ and $h_3$ of opposite sign). This is evident in Figure 15. Similar VPs are observed from the ground (see Figure 11). The mean of the best-fitting Gaussian overestimates the true mean velocity by about 15%, as a result of the VP asymmetry. The mean streaming curves in Figures 13 and 14 have similar shapes. They rise almost linearly out to a characteristic radius which is of the same order as the aperture size, and then remain flat at $\sim 50$ km s$^{-1}$. The circular velocity of the model has a pronounced Keplerian ($v_c \propto r^{-1/2}$) increase close to the black hole. However, only a very minor increase is seen in the predicted mean streaming curve in Figure 13, and no increase at all is seen in Figure 14. The reason for this is that even the smallest aperture available with the HST/FOS is not much smaller than the radius $r_{\rm BH}$ defined in equation (4.2), which for M32 is $\sim 0.1''$.

For $R' = 0.1''$, the mean velocity of the best-fitting Gaussian to the VP is 50 km s$^{-1}$ for the square aperture and 35 km s$^{-1}$ for the circular aperture. If, this close to the centre, mean streaming velocities of this order were actually measured with HST, it would probably provide a strong argument against models without a central black hole. Such models require a large amount of radial motion close to the hole to account for the high central velocity dispersion (vdM94b), and in such models the maximum possible mean streaming is limited (Richstone et al. 1990).

## 5 SUMMARY AND CONCLUSIONS

The contour integral method of Hunter & Qian (1993) can be used to calculate the even part $f_e(E, L_z^2)$ of the DF $f(E, L_z)$ for smooth axisymmetric densities $\rho(R^2, z^2)$ in a potential $\Psi(R^2, z^2)$. Unlike previous methods, the HQ method is applicable in cases where $\rho$ as a function of $\Psi$ and $R^2$ — denoted here by $\tilde{\rho}(\Psi, R^2)$ — is not known explicitly, and this key property finally allows the construction of large classes of realistic axisymmetric galaxy models. We have shown how this can be accomplished for the family of classical spheroids, in which the density distribution is stratified on similar concentric oblate or prolate spheroids with constant axis ratio and has an arbitrary radial profile. In projection, these models have concentric elliptic isophotes with constant ellipticity, and no isophote twist. The "density" $\tilde{\rho}(\Psi, R^2)$ of these models is generally only known implicitly. The HQ method requires evaluation of $\tilde{\rho}$ at complex values of $R^2$ and $\Psi$, and we have described in Section 2 how this analytic continuation can be done numerically. It is then straightforward to evaluate the contour integral for $f_e(E, L_z^2)$.

Our procedure for the calculation of $f(E, L_z)$ applies not only to a single spheroidal component, but also to any combination of them with different axis ratios and density profiles. In particular, it can be applied to the sums of Gaussian density distributions that have been used to represent rather complicated axisymmetric models of realistic galaxies (Monnet, Bacon & Emsellem 1992; Emsellem et al. 1994). In a future paper we shall use our method to provide two-integral DFs for such models and to obtain further insight into their structure and dynamics. It remains to be seen whether $f(E, L_z)$ can be calculated in an analogous way for arbitrary smooth axisymmetric densities that are not of the form $\rho = \rho(m^2)$, for example by direct use of the Poisson integral for the potential (Binney & Tremaine 1987, eq. [2-3]).

In Section 3 we considered a specific set of classical spheroids. These ($\alpha, \beta$)-models have arbitrary axis ratio, and the slopes of the density profile in the inner and outer regions can be chosen independently. The ($\alpha, \beta$)-family contains many popular axisymmetric models as special cases, including the scale-free spheroids in which the density profile is a pure power law. The "density" $\tilde{\rho}(\Psi, R^2)$ of these models is not known explicitly, but it has a simple form, and application of the HQ method is straightforward. We have calculated the resulting DFs for a variety of axis ratios and density profile slopes.

The scale-free spheroids can be compared to the scale-free power-law models of Evans (1993, 1994), for which the potential rather than the density is stratified on similar concentric spheroids. Evans' models have elementary $f_e(E, L_z)$, which lead to elementary and explicit expressions for the observables (Evans & de Zeeuw 1994). However, their density distributions can deviate strongly from a spheroidal shape, and may be peanut-shaped (even though this is less evident in projection). The non-spheroidal shape of these models is reflected in a linear dependence of $f_e(E, L_z)$ on $L_z^2$. By contrast, the DF of the scale-free spheroids has the same energy dependence as the Evans models — which is fixed by the slope of the pure power-law density profile — but the dependence on angular momentum is stronger, so that the high-$L_z$ orbits are more heavily populated. The advantage of the scale-free spheroids presented here is that they have exactly spheroidal density distributions, but this pleasing property comes at a price: the DF and the observables are not elementary functions, and require numerical integrations, which are however straightforward. The scale-free spheroids can be used to approximate the behaviour of the general ($\alpha, \beta$)-models at small and large radii. Their simpler structure speeds up the calculation of $f(E, L_z)$, and hence allows an efficient investigation of parameter space. We have determined in Sections 3.3 and 3.4 the flattenings and density profile slopes for which oblate and prolate cusps have physical, i.e., non-negative, two-integral DFs. We have shown that the physical set of two-integral prolate models is limited in axis ratio, and shrinks even further when a black hole is included in the potential. We also extended the computation of $f_e(E, L_z)$ of the self-consistent scale-free spheroids to the case of scale-free spheroidal densities embedded in Evans' power-law potentials of arbitrary flattening and radial profile. These DFs allow a systematic investigation of the effect of a dark halo on the observed VPs in flattened elliptical galaxies, and hence should be useful for the analysis of kinematic measurements at large radii.

High-resolution ground-based kinematic measurements for the galaxy M32 by vdM94a were interpreted by vdM94b in terms of an $(\alpha,\beta)$-model with $f(E, L_z)$ and a $1.8\times 10^6$ M$_\odot$ central black hole, for which they solved the second and third order moment equations of the collisionless Boltzmann equation. In Section 4 we have computed the exact two-integral DF for this model. We used the HQ method to calculate $f_e$, chose a simple functional form for $f_o$ that fits the observed mean streaming velocities $\langle v_{los}\rangle$, and computed the expected VPs for edge-on observation, taking into account the seeing convolution and spatial binning of the observations. The results confirm that this $f(E, L_z)$ model provides a truly remarkable fit to the available data. In addition, it turns out to have a remarkable property: it is close to being a simple product of a function of energy times a function of the circularity parameter $\eta = L_z/L_c$, with $L_c$ the angular momentum of the circular orbit with energy $E$. Quantitative theoretical work is needed to determine whether this result has any important physical significance.

The success of a two-integral model with a central black hole is no proof that M32 indeed contains such a black hole, as we have not demonstrated that a three-integral axisymmetric (or triaxial, see Emsellem et al. 1993) model without a black hole can be ruled out. Spectroscopic observations with the high spatial resolution of the HST should provide more definite evidence either for or against the presence of a central black hole. We have used our model for M32 to predict what HST should reveal. We calculated the expected VPs for spectroscopic observations with the smallest rectangular ($0.09'' \times 0.09''$) and circular ($D = 0.26''$) apertures available on the HST/FOS. The predicted central Gaussian velocity dispersion is 127 km s$^{-1}$ with the former, and 105 km s$^{-1}$ with the latter aperture. It is not expected that one will be able to measure the expected Keplerian rise of $\langle v_{los}\rangle$ close to the black hole, because its radius of influence is only $\sim 0.1''$. When measured with the available small apertures, the predicted mean streaming motions along the major axis are nearly constant at $\sim 50$ km s$^{-1}$, down to $0.1''$. If such mean streaming motions are indeed measured at $0.1''$ from the centre of M32, then it will be very hard to argue for models without a central black hole. Such models require a strongly radially anisotropic velocity distribution near the centre in order to account for the observed large central velocity dispersion, and hence cannot support large mean streaming motions.

It has finally become practical to calculate $f(E, L_z)$ for realistic axisymmetric galaxy models. We have shown here that one way to do this is to use the HQ method. Other possibilities include the series expansion method of Dehnen & Gerhard (1994) and the grid-based quadratic programming technique of Kuijken (1995). As a result, two-integral axisymmetric models can now replace spherical isotropic models as the standard theoretical templates for a zeroth order comparison to the high quality kinematic observations of flattened elliptical galaxies that are available. The case of M32 shows that $f(E, L_z)$ modelling may already provide a remarkable fit to some galaxies, but it is well-known from modelling based on the Jeans equations that this must be the exception rather than the rule. Application of these improved modelling techniques to elliptical galaxies with more internal structure, such as those with embedded discs, will be quite rewarding.

## ACKNOWLEDGMENTS

The authors are indebted to James Binney, Walter Dehnen and Wyn Evans for useful discussions, and for communication of results prior to publication. This research was supported in part by NSF Grant DMS-9304012. Travel support by the Leids Kerkhoven Bosscha Fonds is acknowledged gratefully.

## APPENDIX A: THREE SPECIAL VELOCITY PROFILES

When the term $-v_{z'}x'\sin i$ in equation (2.15) vanishes, the innermost integral of expression (2.14) for the VP spans an interval of $L_z$ which is symmetric about $L_z = 0$. Consequently only $f_e$ contributes to it. Changing the integration variable to $L_z^2$, the two inner integrals can be written as follows

$$\sigma(\psi,s^2) = 2 \int_{\Psi_\infty}^{\psi} dE \int_0^{2s^2(\psi-E)} \frac{f_e(E,L_z)\,dL_z^2}{\sqrt{L_z^2}\sqrt{2s^2(\psi-E)-L_z^2}}, \quad (A1)$$

where

$$\psi = \Psi(x',y',z') - \tfrac{1}{2}v_{z'}^2,$$
$$s^2 = (-y'\cos i + z'\sin i)^2 + x'^2\cos^2 i. \quad (A2)$$

Equation (A1), which defines the contribution to the even part $VP_e$ of the VP at a spatial point $(x',y',z')$, is equivalent to the fundamental integral equation for two-integral DFs of axisymmetric disc galaxies (e.g. eq. [C1] of HQ). This is not unexpected since in both cases the integrations are carried out over a two-dimensional velocity space. Equation (C2) of HQ transforms the surface "density" of a disc galaxy to a pseudo volume-density of an axisymmetric galaxy. Here we need its inversion to relate $\sigma$ to the "density" $\tilde\rho$. It is

$$\sigma(\psi,s^2) = \frac{1}{\pi\sqrt{2}} \int_{\Psi_\infty}^{\psi} \frac{\tilde\rho_1(\Phi,s^2)}{\sqrt{\psi-\Phi}}\,d\Phi, \quad (A3)$$

in which the subscript 1 denotes the partial derivative with respect to the first argument. Completing the $z'$-integration yields

$$VP_e(v_{z'};x',y') = \frac{1}{\pi\sqrt{2}\Sigma(x',y')} \int_{z_1}^{z_2} dz' \int_{\Psi_\infty}^{\psi} \frac{\tilde\rho_1(\Phi,s^2)}{\sqrt{\psi-\Phi}}\,d\Phi, \quad (A4)$$

always provided $x'v_{z'}\sin i = 0$. We need to have the function $\tilde\rho(\Psi,R^2)$ to use formula (A4) since it is generally difficult to relate $\tilde\rho(\Phi,s^2)$ explicitly to the actual density. When this function is available explicitly, one simply integrates equation (A4). When this function is only know implicitly, one must first evaluate $\tilde\rho(\Phi,s^2)$ by numerical means and then perform numerical integration.

Below we summarize the three cases in which equation (A4) can be used.

### A1 Minor axis ($x' = 0$)

Now equation (A4) gives the VP on the projected minor axis as

$$VP_e(v_{z'};0,y') = \frac{1}{\pi\sqrt{2}\Sigma(x',y')} \int_{z_1}^{z_2} dz' \int_{\Psi_\infty}^{\Psi(0,y',z')-v_{z'}^2/2} \frac{\tilde\rho_1[\Phi,(-y'\cos i + z'\sin i)^2]}{\sqrt{\Psi(0,y',z')-v_{z'}^2/2-\Phi}}\,d\Phi. \quad (A5)$$

The minor axis VP is even, hence the mean line-of-sight velocity there is identically zero, as is physically evident.

### A2 Face–on ($i = 0$)

Now equation (A4) gives the VP of a face-on projection, which is completely even, and is given by

$$VP_e(v_{z'};x',y') = \frac{\sqrt{2}}{\pi\Sigma(x',y')} \int_0^{z_2} dz' \int_{\Psi_\infty}^{\Psi(R'^2,z'^2)-v_{z'}^2/2} \frac{\tilde\rho_1(\Phi,R'^2)}{\sqrt{\Psi(R'^2,z'^2)-v_{z'}^2/2-\Phi}}\,d\Phi, \quad (A6)$$

where $R'^2 = x'^2 + y'^2$.

### A3 Systemic velocity ($v_{z'} = 0$)

In this case equation (A4) gives the density of stars with vanishing line-of-sight velocity in the whole plane of the sky. It is independent of the mean streaming of stars, and is given by

$$VP_e(0;x',y') = \frac{1}{\pi\sqrt{2}\Sigma(x',y')} \int_{-\infty}^{\infty} dz' \int_{\Psi_\infty}^{\Psi(x',y',z')} \frac{\tilde\rho_1[\Phi,x'^2+(-y'\cos i + z'\sin i)^2]}{\sqrt{\Psi(x',y',z')-\Phi}}\,d\Phi. \quad (A7)$$

## APPENDIX B: AUXILIARY RESULTS FOR SCALE–FREE SPHEROIDS

The function $P_{\alpha,q}(\theta)$ that appears in the separated form (3.11) of the potential of the scale-free spheroids is given by

$$P_{\alpha,q} = \frac{1}{\alpha+2} \ln\left[\frac{1}{J_{\alpha,q}} \int_0^\infty \frac{du}{\Delta(u)} \left(\frac{\sin^2\theta}{1+u} + \frac{\cos^2\theta}{q^2+u}\right)^{1+\alpha/2}\right], \quad (B1)$$

when $\alpha \neq -2$, and by its limit

$$P_{-2,q} = \frac{1}{J_{-2,q}} \int_0^\infty \frac{du}{\Delta(u)} \ln(\sin^2\theta + \frac{1+u}{q^2+u}\cos^2\theta), \quad (B2)$$

when $\alpha = -2$. Here we have defined

$$J_{\alpha,q} = \int_0^\infty \frac{du}{\Delta(u)(1+u)^{(\alpha+2)/2}}, \quad (B3)$$

so that $J_{-2,q} = (2/e)\arcsin e$, with $e^2 = 1 - q^2$.

To evaluate the function $\bar{\rho}$ that appears in equation (3.14), we use the explicit formula (3.6) for the density to solve for $\bar{z}^2$ in terms of $\bar{R}^2$ and $\bar{\rho}$, and then substitute the result into the expressions (3.7) and (3.8) for the potential. This gives

$$\int_0^\infty \frac{du}{\Delta(u)} \ln\left[\frac{e^2\zeta u}{1+u} + \frac{1}{\bar{\rho}}\right] = K_q, \qquad (\alpha = -2),$$

$$\int_0^\infty \frac{du}{\Delta(u)(q^2+u)^{(\alpha+2)/2}} \left[\frac{e^2\zeta u}{1+u} + \bar{\rho}^{2/\alpha}\right]^{\frac{\alpha+2}{2}} = J_{\alpha,q}, \quad \text{(B4)}$$

$$(\alpha \neq -2),$$

where $J_{\alpha,q}$ is defined above, and

$$K_q = \int_0^\infty \frac{du}{\Delta(u)} \ln\left(\frac{q^2+u}{1+u}\right). \tag{B5}$$

For given $\alpha$ and $q$, expression (B4) can be solved numerically to find $\bar{\rho}$ for each value of $\zeta$, including complex ones. The value $\bar{\rho}(0)$ can be found explicitly:

$$\bar{\rho}(0) = \begin{cases} \exp(-K_q/I_{-2,q}), & \alpha = -2, \\ (J_{\alpha,q}/I_{\alpha,q})^{\alpha/(\alpha+2)}, & \alpha \neq -2, \end{cases} \tag{B6}$$

where we have written

$$I_{\alpha,q} = \int_0^\infty \frac{du}{\Delta(u)(q^2+u)^{(\alpha+2)/2}}, \tag{B7}$$

so that $I_{-2,q} = J_{-2,q}$. In principle we can also compute the derivatives $\bar{\rho}^{(n)}(0)$ by successive differentiation of the integral equations (B4), and so construct a Taylor series for $\bar{\rho}$. This process provides an approximation of the function $\tilde{\rho}(\Psi, R^2)$, which consists of elementary density components whose DFs can be easily written down (Fricke 1952; Toomre 1982; HQ; Evans 1994). It is hard in practice to carry out this process beyond the first or the second derivative since the differentiations soon lead to lengthy expressions, and its accuracy is as yet unknown. For instance, a linear approximation of $\bar{\rho}(\zeta)$ gives rise to an approximation of $\bar{f}_e(\eta^2)$ which is linear in $\eta^2$, and hence resembles the $\bar{f}_e(\eta^2)$ of Evans's scale-free power-law models. However, for the scale-free spheroids this approximation is not accurate unless they are very nearly spherical.

We can compute $\bar{f}_e(\eta^2)$ accurately by the contour integral (2.2). By substitution of equations (3.14) into equation (2.2), carrying out partial derivatives, use of equation (3.15), transformation to the scaled variables (3.16), and use of certain simple transformations, we arrive at integrals for $\bar{f}_e(\eta^2)$. We need to distinguish three cases: $\alpha = -2$, $-3 < \alpha < -2$ and $-2 < \alpha < 0$. The result for the $\alpha = -2$ case is

$$\bar{f}_e(\eta^2) = \frac{1}{2\pi^2 i} \int_{-\infty}^{(1+)} \frac{\exp(t-1)\,dt}{\sqrt{t}} H(X), \tag{B8}$$

where $X = \eta^2 \exp(t-1)/t$, and the function $H$ is defined as

$$H(X) = -\frac{\alpha}{2}\bar{\rho}(X) + (2 - \frac{\alpha}{2})X\bar{\rho}'(X) + X^2\bar{\rho}''(X). \tag{B9}$$

For $-3 < \alpha < -2$ we have

$$\bar{f}_e(\eta^2) = \frac{(t_0-1)^{-1/2}}{2\pi^2 i} \int_0^{(t_0+)} \left(\frac{t_0}{t}\right)^{1/(1-t_0)} \frac{dt}{\sqrt{t-1}} H(Y), \tag{B10}$$

where $t_0 = 2/(\alpha+4)$, and $Y = \eta^2 (t/t_0)^{t_0/(t_0-1)}(t_0-1)/(t-1)$. When $\alpha > -2$,

$$\bar{f}_e(\eta^2) = \frac{(1-t_0)^{-1/2}}{2\pi^2 i} \int_{-\infty}^{(-t_0+)} \left(-\frac{t_0}{t}\right)^{1/(1-t_0)} \frac{dt}{\sqrt{t+1}} H(Z), \tag{B11}$$

where $Z = \eta^2(-t_0/t)^{t_0/(1-t_0)}(1-t_0)/(1+t)$. When $\eta = 0$, $X = Y = Z = 0$. The reduced contour integrals in equations (B8), (B10) and (B11) can then be evaluated by wrapping the contours tightly around respective branch cuts to convert them to elementary real integrals. The results are

$$\bar{f}_e(0) = \frac{\bar{\rho}(0)}{\pi^{3/2} \exp(1)}, \qquad \alpha = -2, \tag{B12}$$

and

$$\bar{f}_e(0) = \frac{(-\alpha)t_0^{\frac{1}{1-t_0}} \bar{\rho}(0)}{2\pi^2 \sqrt{|t_0-1|}} \times \begin{cases} B(\frac{1}{2}, \frac{t_0}{t_0-1}), & \alpha < -2, \\ B(\frac{1}{2}, \frac{1}{1-t_0} - \frac{1}{2}), & \alpha > -2, \end{cases} \tag{B13}$$

and $\bar{\rho}(0)$ is given in equation (B6). In the numerical evaluation of integrals (B8) and (B11) we have employed the new integration variables $s = \exp(t-1)$ and $s = -1/t$, respectively, so the resulting integrals are along finite paths.

The case of a density $\rho_0(m/b)^\alpha$ in a non-self-consistent power-law potential $\Psi_d$ of the form (3.25), discussed in Section 3.4, leads to DFs of the form (3.28). The functions $\bar{f}_{\alpha,\gamma}(e_d^2\eta^2)$ that appear there can be computed by integrals of the form (B8), (B10) and (B11), for cases with $\gamma = 0$, $\gamma < 0$ and $\gamma > 0$, respectively, with the following modifications. We replace $\alpha + 2$ by $\gamma$ in the definition of $t_0$. The function $H$ now is

$$H(X) = \frac{\alpha}{2}\left[\left(\frac{\gamma}{2}-1\right)e_d^2 X + \frac{\alpha}{2} - \frac{\gamma}{2}\right](1-e_d^2 X)^{\frac{\alpha}{2}-2}. \tag{B14}$$

The exponential term in the integral (B8) for the case $\gamma = 0$ is replaced by $\exp[-\alpha(t-1)/2]$. The power terms in the integrals (B10) and (B11) for the cases $\gamma \neq 0$ have exponent $2-(\alpha/\gamma)$. However, the major difference is that the function $\bar{\rho}(\zeta)$ is now equal to $(1 - e_d^2\zeta)^{\alpha/2}$, and hence is explicit, so that the integrations can be evaluated numerically in a straightforward manner.

# APPENDIX C: VELOCITY DISPERSIONS IN THE SCALE-FREE SPHEROIDS

The explicit solution of the Jeans equations for an $f(E, L_z)$ axisymmetric model is (Hunter 1977)

$$\rho\sigma_R^2 = \rho\sigma_z^2 = \int_{\Psi_\infty}^{\Psi(R^2, z^2)} \tilde{\rho}(\Phi, R^2)\, d\Phi,$$

$$\rho\langle v_\phi^2\rangle = \int_{\Psi_\infty}^{\Psi(R^2, z^2)} \frac{\partial}{\partial R}[R\tilde{\rho}(\Phi, R^2)]\, d\Phi. \tag{C1}$$

**Figure C1.** Dynamical quantities in the meridional plane for $f(E, L_z)$ singular isothermal spheroids with four different axis ratios, as functions of the latitudinal angle $\theta$ (in degrees). The value $\theta = 0°$ corresponds to the symmetry axis, the value $\theta = 90°$ to the equatorial plane. Solid curves: $\sigma_R = \sigma_z$; Dashed curves: $\langle v_\phi^2 \rangle^{1/2}$. The quantities are given in units of $V_0$, the circular velocity in the equatorial plane. This figure was adapted from van der Marel (1994e).

The mean streaming motions $\langle v_R \rangle$ and $\langle v_z \rangle$ vanish in an axisymmetric model, so that the second moments $\langle v_R^2 \rangle$ and $\langle v_z^2 \rangle$ are equal to the dispersions $\sigma_R^2$ and $\sigma_z^2$.

For the scale-free spheroids, the function $\tilde{\rho}(\Phi, R^2)$ is given in terms of the single-variable function $\bar{\rho}$. A simple relation between the two identical dispersions $\sigma_R^2$, $\sigma_z^2$ and the second moment $\langle v_\phi^2 \rangle$ can then be established by a judicious combination of the equations above such that the resulting integrand is a total derivative. We find

$$\langle v_\phi^2 \rangle - (2\alpha + 3)\sigma_R^2 = V_0^2 - (\alpha + 2)\Psi, \quad (C2)$$

where $\Psi$ is the potential given by equation (3.8) (for $\alpha \neq -2$) or (3.7) (for $\alpha = -2$) and the constant $V_0$ is given in equation (3.10). This relation is valid locally, i.e., at each point $(R, z)$. Similar relations can be derived for the higher order even moments. We note that we must have $\alpha < -1$ for $\langle v_\phi^2 \rangle$ and $\sigma_R^2$ to be finite everywhere.

We now restrict our attention to the singular isothermal spheroids (i.e., $\alpha = -2$). The velocity dispersions are then elementary. The equation (C1) for $\sigma_R^2 = \sigma_z^2$ can be written as

$$\rho \sigma_R^2 = \rho \sigma_z^2 = -\int_z^\infty \rho(R^2, z'^2) \frac{\partial \Psi}{\partial z}(R^2, z'^2)\, dz'. \quad (C3)$$

Differentiating equation (3.7) gives

$$\frac{\partial \Psi}{\partial z} = -\frac{V_0^2}{r \arcsin e} \arctan \frac{ez}{qr}. \quad (C4)$$

Upon substitution of

$$t = \arctan \frac{ez'}{q\sqrt{R^2 + z'^2}}, \quad (C5)$$

the integral in equation (C3) can be carried out, with result:

$$\sigma_R^2 = \frac{V_0^2 q}{2e \arcsin e}\left(1 + \frac{\cot^2 \theta}{q^2}\right) \times \left[\left(\arctan \frac{e}{q}\right)^2 - \left(\arctan \frac{e \cos \theta}{q}\right)^2\right]. \quad (C6)$$

The remaining dispersion $\langle v_\phi^2 \rangle$ now follows trivially from equation (C2), which for $\alpha = -2$ reduces to

$$\langle v_\phi^2 \rangle + \sigma_R^2 = V_0^2. \quad (C7)$$

The constant $V_0$ is equal to the circular velocity in the equatorial plane for these models. Both $\sigma_R^2$ and $\langle v_\phi^2 \rangle$ are independent of radius, but do depend on the polar angle $\theta$ defined by $R = r \sin\theta$ and $z = r \cos\theta$. The total second moment parallel to the equatorial plane, $\langle v_\phi^2 \rangle + \sigma_R^2$, is independent of $\theta$, and is always equal to $V_0^2$. On the $z$-axis ($\theta = 0$) we have $\sigma_R^2 = \sigma_z^2 = \langle v_\phi^2 \rangle = V_0^2/2$. The second moments are non-negative when $0 \leq q \leq 3.46717$. This does not imply that $f_e \geq 0$ for all these models. The analysis in Section 3.2 shows that $f_e \geq 0$ only when $q \leq 1.3903$. Figure C1 shows the dynamical quantities of the singular isothermal spheroids for various axis ratios, in units of $V_0$. In the equatorial plane $\langle v_\phi^2 \rangle > \sigma_R^2$ for oblate models, and $\langle v_\phi^2 \rangle < \sigma_R^2$ for prolate models.

## APPENDIX D: SEEING CONVOLUTION

Any observed quantity is a line-of-sight projected quantity, averaged over some finite pixels on a detector. For ground-based observations we also have to take into account the effect of atmospheric seeing, which convolves the projected properties of the galaxy with a point spread function. This PSF is often taken as Gaussian or as the sum of Gaussians. Here we restrict ourselves to Gaussians that are circular on the plane of the sky.

Consider a rectangular pixel **R** of size $2l \times 2w$, whose centre is at the point $(x_0', y_0')$ and whose axes make an arbitrary angle $\Theta_0$ with the $x'$-axis on the sky. We define a new $(\tilde{x}, \tilde{y})$-coordinate system with its centre at $(x_0', y_0')$ and with axes parallel to the sides of the pixel (Figure D1). The associated transformation is then

$$\begin{aligned} x' &= x_0' + \tilde{x}\cos\Theta_0 - \tilde{y}\sin\Theta_0, \\ y' &= y_0' + \tilde{x}\sin\Theta_0 + \tilde{y}\cos\Theta_0. \end{aligned} \quad (D1)$$

Let $S(x', y')$ be a function that depends linearly on the luminosity density of the galaxy, e.g., $\Sigma(x', y') \times \mathrm{VP}(v_{z'}; x', y')$, as defined by equation (2.13). Assuming the PSF to be a single normalized Gaussian, $A \exp(-R'^2/2\sigma^2)$, the seeing convolved function $S_s$ satisfies:

$$S_s(x_0' + x', y_0' + y') = A \int_{-\infty}^{\infty}\int_{-\infty}^{\infty} dx\, dy \quad (D2)$$
$$S(x_0' + x, y_0' + y) e^{-[(x-x')^2 + (y-y')^2]/2\sigma^2}.$$

The observable $S_o$ at the position $(x_0', y_0')$ is:

$$S_o(x_0', y_0') = \frac{1}{4lw}\int\!\!\int_{\mathbf{R}} S_s(x', y')dx'dy' = \frac{1}{4lw}\int_{-l}^{l}\!\!\int_{-w}^{w} d\tilde{x}\,d\tilde{y} \quad (D3)$$
$$S_s(x_0' + \tilde{x}\cos\Theta_0 - \tilde{y}\sin\Theta_0, y_0' + \tilde{x}\sin\Theta_0 + \tilde{y}\cos\Theta_0).$$

**Figure D1.** The coordinate system used for the seeing convolution of observations with finite pixels of size $2l \times 2w$.

Substituting equation (D2) into equation (D3) and exchanging the order of integrations, we find that the $\tilde{x}$- and $\tilde{y}$-integrations can be carried out to give expressions in terms of error functions. A final use of polar coordinates defined by $x = r\cos(\theta + \Theta_0)$ and $y = r\sin(\theta + \Theta_0)$, which contains a rotation in the $(x, y)$-coordinates, yields

$$S_\circ(x_0', y_0') = \frac{1}{4lw} \int_0^\infty r\,dr \int_0^{2\pi} d\theta\, K(r,\theta) \quad\quad (D4)$$
$$S[x_0' + r\cos(\theta + \Theta_0), y_0' + r\sin(\theta + \Theta_0)],$$

where

$$K(r,\theta) = \frac{A\sigma^2\pi}{2} \left[\mathrm{erf}\left(\frac{l + r\cos\theta}{\sqrt{2}\sigma}\right) - \mathrm{erf}\left(\frac{-l + r\cos\theta}{\sqrt{2}\sigma}\right)\right] \\ \times \left[\mathrm{erf}\left(\frac{w + r\sin\theta}{\sqrt{2}\sigma}\right) - \mathrm{erf}\left(\frac{-w + r\sin\theta}{\sqrt{2}\sigma}\right)\right]. \quad (D5)$$

When the PSF is a sum of Gaussians, the kernel $K$ is then also a sum of expressions given by equation (D5). Equation (D4) therefore allows us to combine seeing convolution and pixel averaging in one step. The error functions in the kernel $K$ can be computed using efficient algorithms. We note that for the given size of pixel and PSF, $K$ is independent of the position $(x_0', y_0')$ on the sky, hence can be computed beforehand on a grid in $(r, \theta)$ space.